\documentclass[showpacs,preprintnumbers,amsmath,amssymb]{revtex4}
\newcommand{\be}{\begin{equation}}
\newcommand{\ee}{\end{equation}}
\newcommand{\bea}{\begin{eqnarray}}
\newcommand{\eea}{\end{eqnarray}}
\renewcommand{\vec}[1]{\mbox{\boldmath$#1$}}
\begin{document}
\title{Massive CP$^1$ theory from a microscopic model for doped antiferromagnets}
\author{J.\ Falb, 
and A.\ Muramatsu }
\affiliation{Institut f\"ur Theoretische Physik III, Universit\"at Stuttgart,
Pfaffenwaldring 57, D-70550 Stuttgart, Federal Republic of Germany}
\begin{abstract}
A path-integral for the $t$-$J$ model in two dimensions is constructed based 
on Dirac quantization, with an action found originally by Wiegmann
(Phys.\ Rev.\ Lett.\ {\bf 60}, 821 (1988); Nucl.\ Phys.\ B323, 311 (1989)). 
Concentrating on the low doping limit, we assume short range antiferromagnetic
order of the spin degrees of freedom. Going over to a local spin quantization 
axis of the dopant fermions, that follows the spin degree of freedom, 
staggered CP$^1$ fields result and the constraint against double occupancy 
can be resolved. The staggered CP$^1$ fields are split into slow and fast 
modes, such that after a gradient expansion, and after integrating out the 
fast modes and the dopant fermions, a CP$^1$ field-theory 
with a massive gauge field is obtained that
describes generically incommensurate coplanar magnetic structures, as 
discussed previously in the context of frustrated quantum antiferromagnets. 
Hence, the possibility of deconfined spinons is opened by doping a collinear
antiferromagnet.  
\end{abstract}
\maketitle

\section{Introduction}
High temperature superconductivity (HTS) remains an unresolved problem
in spite of an enormous research effort over more than twenty years  
\cite{science06}. On the theoretical side, a number of phenomenological 
theories were proposed \cite{sachdev03,kivelson03,demler04} that may describe 
certain aspects of the experimental findings, however, they rely on 
fundamentally different assumptions, so that it is, at this point, difficult 
to assess their validity. Also approaches based on microscopic models like the 
$t$-$J$ one were advanced \cite{lee06}, based on the so-called slave boson 
formulation, where electrons are split into separated spin (spinon)
and charge (holons) degrees of freedom. Such a separation introduces a
local gauge invariance, that renders mean-field like approximations 
particularly troublesome. A common feature of the above mentioned approaches 
is the difficulty to connect in a controlled way the states proposed for the 
doped case to the undoped one. 

In fact, the theoretical description of the evolution from a N\'eel state
in the parent compounds towards the state of a doped quantum antiferromagnet, 
lies at the heart of a theoretical understanding of HTS.
A phenomenological description of doped antiferromagnets was first given
by Shraiman and Siggia in a series of seminal papers 
\cite{shraiman88b,shraiman89,shraiman90,shraiman92}. The picture emerging from 
a semiclassical treatment of mobile holes in an antiferromagnet corresponds
to a coplanar twist of the spin background that gives rise to a dipolar 
field centered on the dopant holes. Similar coplanar structures are also 
expected in frustrated quantum antiferromagnets 
\cite{azaria90,azaria93,chubukov94}.

The field-theoretic treatment of frustrated quantum antiferromagnets showed on
the one-hand, that an O(4) symmetry is dynamically generated, and in two
dimensions at temperature $T=0$ long-range order can set in either
through a first 
order transition or a second order one with exponents corresponding to 
an O(4) non-linear $\sigma$ model \cite{azaria90,azaria93}. On the other 
hand, when the effective theory is formulated in terms of CP$^1$ fields,
the presence of deconfined spinons can be induced from the fact that the 
corresponding gauge field becomes massive 
\cite{chubukov94,azaria95,chubukov96}.
Yet, an explicit connection between doped antiferromagnets and the effective 
theories for frustrated quantum antiferromagnets is missing. 

The experimental observation of incommensurate peaks in neutron scattering 
experiments on La$_{2-x}$Sr$_x$CuO$_4$ (LSCO) \cite{matsuda02} and on 
YBa$_2$Cu$_3$O$_7$ (YBCO) \cite{bourges00,mook02,hayden04,hinkov04} revived 
the interest 
on coplanar magnetic configurations in cuprates. A number of theoretical 
works \cite{juricic04,sushkov05,lindgard05,juricic06,luescher07}
focused on material specific descriptions of the magnetic structures
at low doping, mostly based on the phenomenology developed by Shraiman and 
Siggia. Although these works were very successful in interpreting the 
experimental results, demonstrating that the used phenomenology describes 
very well the generic features at low doping, they did not make an explicit
connection to a microscopic model.

An alternative way to describe the low energy physics of doped 
antiferromagnets is the development of an effective field-theory
based on the symmetries of the system (global SU(2) and U(1) in the case
of a doped quantum antiferromagnet) and the possible spontaneously broken
symmetry phases. Such a path, reminiscent of chiral perturbation theory
was followed starting with the Hubbard model, as a representative one
for doped antiferromagnets \cite{kaempfer05}. At the moment the predictive 
power of the derived effective action is not clear.

The most direct approach for treating a microscopic model like the 
Hubbard \cite{anderson87} or $t$-$J$ \cite{zhang88} models encountered until 
now a number of difficulties. While on the one hand, a mean-field treatment of 
the $t$-$J$ model based on the slave boson approach \cite{lee06} led to a 
qualitative understanding of various experimental results, like the existence 
of a pseudogap, it is difficult to assess their reliability due to the 
uncontrolled nature of the mean-field approximation. The inclusion of 
fluctuations of the gauge fields remains until now difficult, since in this 
case their coupling to the matter fields is strong. Although their treatment 
led to a qualitative description of the doped phase, still progress would be
desirable on a more quantitative basis \cite{lee06}. On the other hand, 
numerically exact results could be only obtained by diagonalization in rather 
small clusters \cite{dagotto94} so that their interpretation remains 
inconclusive when contrasted to the low energy behavior in HTS. For larger
sizes, variational Monte Carlo techniques indicate that the $t$-$J$ model 
supports
superconductivity \cite{sorella02}. Still, in spite of the accuracy of the 
method, further insight in the system is still missing. Unfortunately, large
scale quantum Monte Carlo simulations are hindered by the so-called minus sign 
problem that affects the simulation of doped antiferromagnets.

In view of the described situation, a controlled analytical treatment of a 
microscopic model is desirable, in particular with the possibility of 
examining the low doping regime, such that the change from an antiferromagnet 
with long-range order at zero doping to a doped situation can be followed in
detail. A first step in that direction was made by Wiegmann \cite{wiegmann88},
who obtained an action for the $t$-$J$ model based on the coherent-states 
method
\cite{wiegmann89}. The same action was found by using a supersymetric version
of the Faddeev-Jackiw symplectic formalism \cite{faddeev88} applied to the 
$t$-$J$ model \cite{foussats99}, with explicit expressions for the measure of 
the 
path integral. 

In the present work we use alternatively the well known procedure of Dirac 
quantization for constrained systems \cite{weinberg05V1} to set up the path
integral, recovering the results in Ref.\ \cite{foussats99}, as shown in Sec.\
\ref{DiracQ}. There we shortly review the $t$-$J$ model and its 
representation in terms of Hubbard $X$-operators 
\cite{hubbard63,hubbard64a,hubbard64b}, that fulfill the graded Lie algebra 
Spl(2,1) \cite{wiegmann88,wiegmann89,scheunert77}. Introducing a series of
primary constraints and on the basis of the action proposed by Wiegmann
\cite{wiegmann88,wiegmann89}, all the constraints of the theory are determined.
It turns out that only second class constraints appear, such that they can be 
solved by inverting the matrix of constraints, and the Dirac brackets reproduce
the algebra of the $X$-operators. On this basis, the path integral can be set 
up. Passing from the representation by $X$-fields to real vector fields for 
the spin degrees of freedom and Grassmann variables for the dopant holes 
\cite{wiegmann88,foussats99}, we arrive at the action that will be the 
starting point for a long-wavelength expansion. In order to proceed further, 
we restrict ourselves to the low doping limit and neglect terms quadratic in 
the density of dopant holes.

The long-wavelength expansion is performed in Sec.\ 
\ref{StaggCP1}, where a staggered CP$^1$ representation is introduced for the
spin-fields. Using the CP$^1$ representation it is possible to resolve 
exactly the constraint against double occupancy \cite{wiegmann88}, that is in 
general the 
stumbling block for a controlled treatment of the model. 
Slow and fast modes of the CP$^1$ fields are identified in 
the same spirit as done for vector fields in quantum antiferromagnets
\cite{fradkin91}. The effective action for the magnetic degrees of freedom 
is reached after integrating out the fast CP$^1$ modes and the fermions.
The resulting effective field-theory corresponds to a CP$^1$ model with 
a massive gauge field, as was generally discussed in the context of frustrated
quantum antiferromagnets \cite{chubukov94,azaria95,chubukov96}. In the 
present work,
however, we obtained the explicit doping dependence of the coupling constants.
In Sec.\ \ref{conclusion} we discuss the obtained results that open the 
possibility of having deconfined spinons by doping.
Some intermediate results are presented in the appendices that may be helpful 
for readers interested in reproducing our results.  

\section{Dirac quantization of the 
$t$-$J$ model \label{DiracQ}} 
We introduce first the $t$-$J$ model and its representation in terms of
so-called $X$-operators that operate only in the subspace without 
doubly occupancy. After discussing shortly the algebra they fulfill,
we delineate the procedure of Dirac quantization.

\subsection{The $t$-$J$ model and $X$-operators\label{XOperators}}
The $t$-$J$ model is defined by the following Hamiltonian in second 
quantization:
\be
\label{hamilt}
H_{t-J} = -t \sum\limits_{<i,j> \atop \sigma}  
{\tilde c}^\dagger_{i\sigma} {\tilde c}^{}_{j\sigma}  
+\frac{J}{2} \sum\limits_{<i,j>} \left({\vec S}_i \cdot {\vec S}_j - 
\frac{1}{4} {\tilde n}_i {\tilde n}_j \right) 
- \mu \sum_i {\tilde n}_i
\; , 
\ee
with ${\tilde c}^{\dag}_{i\sigma} = (1-n^{}_{i-\sigma})c^{\dag}_{i\sigma}$,
${\tilde n}_i = \sum_\sigma {\tilde c}^{\dag}_{i \sigma} 
{\tilde c}^{}_{i \sigma}$,
${\vec S}_i = \sum\limits_{\sigma,\sigma'} 
c^{\dag}_{i\sigma} {\vec \sigma}_{\sigma \sigma'} c^{}_{i\sigma'}$, where the 
operators $c^\dagger_{i\sigma}$ and $c^{}_{i\sigma}$ denote canonical 
creation and annihilation operators, respectively, for electrons at site $i$ 
and spin indices $\sigma = \pm$. The operators ${\tilde c}^{\dag}_{i\sigma}$
and ${\tilde c}^{}_{i\sigma}$ project out doubly occupied states. 
Then, $t$ gives the hopping amplitude, $J$ is the antiferromagnetic exchange 
coupling and $\mu$ the chemical potential. The symbol $<i,j>$ restricts the
sums to nearest neighbors. The
Hamiltonian (\ref{hamilt}) was obtained from a multiband model 
\cite{zhang88} and represents the minimal model for cuprates.

Introducing so-called $X$-operators \cite{hubbard63,hubbard64a,hubbard64b}
defined as
\bea
{\hat X}_i^{\alpha \beta} & = & \mid \alpha_i > < \beta_i \mid
\; ,
\eea
with $\alpha_i = 0,\sigma$ for site $i$, the Hamiltonian becomes a bilinear
form in such operators.
\bea
\label{hamilt'}
H_{t-J} & = & -t \sum\limits_{<i,j> \atop \sigma} 
{\hat X}_i^{\sigma 0} {\hat X}_j^{0 \sigma} 
+ \frac{J}{4} \sum_{<i,j> \atop \sigma, \bar{\sigma}}
\left( {\hat X}_i^{\sigma \bar{\sigma}} {\hat X}_j^{\bar{\sigma} \sigma}
-{\hat X}_i^{\sigma \sigma} {\hat X}_j^{\bar{\sigma} \bar{\sigma}} \right)
- \mu \sum_{i,\sigma} {\hat X}_i^{\sigma 0} {\hat X}_i^{0 \sigma}
\; .
\eea
The $X$-operators fulfill the following graded algebra
\bea
\label{Xalgebra}
\left[ {\hat X}_i^{\alpha\beta} , {\hat X}_j^{\gamma\delta} \right]_\pm =
\delta_{ij} \left( \delta^{\beta \gamma} {\hat X}_i^{\alpha\delta} \pm
\delta^{\alpha \delta} {\hat X}_i^{\gamma\beta} \right) \; ,
\eea
with $-$ ($+$) corresponding to a commutator (anticommutator). 
Anticommutation relations appear only when both operators are fermionic.
Furthermore, we have the completeness condition
\bea
\label{XComplete}
\sum_\alpha {\hat X}^{\alpha \alpha}_i = \vec 1 \; .
\eea 

A further insight into the graded algebra above can be gained by considering
the commutation and anticommutation relations of the even (bosonic) and odd
(fermionic) parts. Denoting the even generators by $Q_m$, $m=1,2,3$ and $B$,
and the odd ones by $U_i$, $i=1,\dots,4$, the commutation relations of the 
Spl(2,1) algebra are \cite{scheunert77}
\bea
\label{CommutationSpl21}
\left[ Q_m , Q_n \right] & = & i \varepsilon_{mnp} Q_p \; , \quad
\left[ Q_m , B \right] = 0 \; ,
\nonumber \\
\left[ Q_m , U_\alpha \right] & = & \frac{1}{2} {\hat \sigma}^m_{\beta \alpha}
U_\beta \; , \quad \;  
\left[ B , U_\alpha \right] = 
\frac{1}{2} {\hat \epsilon}_{\beta \alpha} U_\beta \; ,
\nonumber \\
\left\{ U_\alpha , U_\beta \right\} & = &
\left( \hat C {\hat \sigma}^m \right)_{\alpha \beta} Q_m -
\left(\hat C \hat \epsilon \right)_{\alpha \beta} B \; , 
\eea
where the $4\times 4$ matrices ${\hat \sigma}^m$, $\hat C$, and $\hat \epsilon$
are defined as follows,
\bea
{\hat \sigma}^m & = & \left(
\begin{array}{cc}
\sigma^m & 0 \\
0 & \sigma^m 
\end{array}
\right) \; , \quad
\hat C = \left(
\begin{array}{cc}
0 & C \\
C & 0 
\end{array}
\right) \; , \quad
\hat \epsilon = \left(
\begin{array}{rr}
\vec 1 & 0 \\
0 & -\vec 1
\end{array}
\right) \; ,
\eea
with $\sigma^m$, $m=1,2,3$ the Pauli matrices and $C=i \tau^2$. In the 
irreducible representation corresponding to the $X$-operators, the generators
above look as follows. For the even sector we have
\bea
Q_1 & = & \frac{1}{2} \left( {\hat X}^{+-} + {\hat X}^{-+} \right) \; ,
\quad
Q_2 = -\frac{i}{2} \left( {\hat X}^{+-} - {\hat X}^{-+} \right) \; ,
\nonumber \\
Q_3 & = & \frac{1}{2} \left( {\hat X}^{++} - {\hat X}^{--} \right) \; ,
\quad
B = \frac{1}{2} \left( {\hat X}^{++} + {\hat X}^{--} \right) - \vec 1 \, .
\eea
The first three generators are those of SU(2) while the last one corresponds 
essentially to particle number. Here and in the following we eliminate 
${\hat X}^{00}$ using the completeness relation (\ref{XComplete}).
For the odd sector we have
\bea
U_1 & = & {\hat X}^{+0} \; , \quad U_2 =  {\hat X}^{-0} \; , \quad
U_3 = {\hat X}^{0-} \; , \quad U_4 = - {\hat X}^{0+} \; .
\eea
As shown in Ref.\ \cite{scheunert77}, the Casimir operator quadratic 
in the generators is given by 
\bea
\label{CasimirOperatorK2}
K_2 & = & {\vec Q}^2 - B^2 + \frac{1}{2} U {\hat C} U \; ,
\eea
that in terms of the $X$-operators looks as follows.
\bea
\label{CasimirOperatorK2'}
K_2  & = & \frac{1}{2} \left({\hat X}^{+-} {\hat X}^{-+}
+  {\hat X}^{-+} {\hat X}^{+-} \right)
+ \frac{1}{4} \left( {\hat X}^{++} - {\hat X}^{--} \right)^2
- \left[ \frac{1}{2} \left( {\hat X}^{++} 
+ {\hat X}^{--} \right) - \vec 1 \right]^2
\nonumber \\ & &
+ \frac{1}{2} 
\left(- {\hat X}^{+0}  {\hat X}^{0+} - {\hat X}^{-0}  {\hat X}^{0-}
+ {\hat X}^{0-}  {\hat X}^{-0} + {\hat X}^{0+}  {\hat X}^{+0} \right)
\; .
\eea
Then, it is easily seen, that this Casimir operator has eigenvalue zero
in the present irreducible representation.

\subsection{Dirac quantization for the $t$-$J$
model}
In the following we consider a {\em classical} system with the 
Lagrangian found by Wiegmann \cite{wiegmann88,wiegmann89}, expressed in 
terms of $X$-fields \cite{foussats99}, with, as usual, complex fields 
corresponding to bosonic operators in Sec.\ \ref{XOperators}, and
Grassmann fields for fermionic ones. 
\bea
\label{Lagrangean}
L \left(X, \dot{X} \right) & = & 
-i \sum_i 
\frac{\left(1+\rho_i \right) u_i -1}
{\left(2-v_i\right)^2 - 4 \rho_i -u_i^2}
\left(X_i^{-+} {\dot X}_i^{+-} - X_i^{+-} {\dot X}_i^{-+} \right)
+ \frac{i}{2} \sum_{i, \sigma} 
\left( X_i^{\sigma 0} {\dot X}_i^{0 \sigma} 
+  X_i^{0 \sigma} {\dot X}_i^{\sigma 0} \right) - H\left(X\right) ,
\eea 
where the following definitions were introduced
\bea
\label{DefsVeariables}
\rho_i \equiv X_i^{0 +} X_i^{+ 0} +  X_i^{0 -} X_i^{- 0}  \; , \quad
u_i \equiv X_i^{+ +} - X_i^{--} \; , \quad
v_i \equiv X_i^{+ +} + X_i^{--} \; .
\eea
Furthermore,we take into account the set of primary constraints found in the 
supersymetric extension of the symplectic formalism introduced by Faddeev and 
Jackiw \cite{foussats99}, where we omit the site indices.
\bea
\label{Constraint(1)}
\phi^{(1)} & = & X^{++} + X^{--} + \rho -1 \; ,
\nonumber \\
\phi^{(2)} & = & X^{+-} X^{-+} + \frac{1}{4} u^2 
- \left(1 - \frac{1}{2} v \right)^2 + \rho \; ,
\nonumber \\
\phi^{(7)} & = & X^{00} - \left(
X^{0+} X^{+0} + X^{0-} X^{-0} \right) \; ,
\nonumber \\
\phi^{(9)} & = & \frac{X^{0+} X^{+-}}{ X^{++}} - X^{0-} \; ,
\nonumber \\ 
\phi^{(10)} & = & X^{+0} X^{-+} - X^{-0} X^{++}
\; .
\eea
The constraints are imposed by setting $\phi^{(a)} = 0$ and the order 
of the labels $a$ is such that the supermatrix of constraints has a 
normal form \cite{gitman90,cornwell89V3}. The special choice of $\phi^{(9)}$
was made in order to have a simple expression for the measure of the 
path integral obtained at the end. The constraint $\phi^{(1)}$ (together
with $\phi^{(7)}$) corresponds to the completeness relation 
(\ref{XComplete}), while $\phi^{(2)}$ comes from the Casimir operator
(\ref{CasimirOperatorK2'}) having eigenvalue zero. Furthermore, $\phi^{(7)}$
relates the empty sites with fermionic holes, and the rest of the constraints
above define new product rules instead of the ones obeyed by the operators
in Sec.\ \ref{XOperators} \cite{foussats99}. 

There are other nine primary constraints resulting from considering the 
canonical momenta
\bea
\label{CanonicalMomenta}
\Pi^{\alpha \beta} & = & 
\frac{\partial_r L}{\partial {\dot X}^{\alpha \beta}} \; ,
\eea
where $\partial_r L/\partial {\dot X}^{\alpha \beta}$ refers to the right
derivative, the $X^{\alpha \beta}$ and its derivatives being elements of the 
Berezin algebra \cite{gitman90} (see Appendix 
\ref{ListOfConstraints}). 
The additional nine constraints are listed in Appendix \ref{ListOfConstraints}.

Once the constraints are determined, a Hamiltonian 
\bea
H = \sum_{\alpha \beta} \Pi^{\alpha \beta} {\dot X}^{\alpha \beta} - L
\eea
can be in principle obtained, where we have to take into account the 
constraints. This can be done introducing Lagrange multipliers 
$\lambda$ that enter in
the equations of motion for any observable $f(X,\Pi)$. In particular,
if the primary constraints are required to apply at any time, we should
require that 
\bea
\label{MotionOfConstraints}
{\dot \phi}^{(a)} & = & \left\{H,\phi^{(a)}\right\} 
+ \lambda_b \left\{\phi^{(b)},\phi^{(a)} \right\} = 0 \; ,
\eea
where the curly brackets denote now Poisson brackets and summation over
repeated indices is assumed. It is understood that the constraints are applied 
after the derivatives in the Poisson brackets are calculated. 
For the Poisson 
brackets among the constraints we have the general form \cite{gitman90} 
\bea
\label{SuperMatrixConst}
\left\{ \phi^{(a)} , \phi^{(b)} \right\} & = &
\sum_{k \atop \alpha , \beta}
\left[ 
\frac{\partial_r \phi^{(a)}}{\partial X_k^{\alpha \beta}}\,
\frac{\partial_\ell \phi^{(b)}}{\partial \Pi_k^{\alpha \beta}} -
(-1)^{P_{\phi^{(a)}} P_{\phi^{(b)}}} 
\frac{\partial_r \phi^{(b)}}{\partial X_k^{\alpha \beta}}\,
\frac{\partial_\ell \phi^{(a)}}{\partial \Pi_i^{\alpha \beta}} 
\right] \; ,
\eea 
where $P_{\phi^{(a)}}$ is the parity of $\phi^{(a)}$, and 
$\partial_\ell \phi^{(b)}/\partial \Pi_k^{\alpha \beta}$ refers to the
left derivative \cite{gitman90}, that is defined in a similar way as the
right derivative in (\ref{CanonicalMomenta}) (see Appendix 
\ref{ListOfConstraints}). 
If the matrix of constraints $\left\{\phi^{(a)},\phi^{(b)} \right\}$ is not 
singular, then the constraints are second class \cite{weinberg05V1}.

In our case, the matrix of Poisson brackets can be written as a 
supermatrix of the form
\bea
\label{MatrixOfConstraints}
\left\{\phi^{(a)},\phi^{(b)}\right\} & = & \left[
\begin{array}{cc}
\vec A & \vec B \\
\vec C & \vec D
\end{array}
\right] \; ,
\eea
where $\vec A$ is an $8\times 8$, $\vec B$ an $8\times 6$, $\vec C$ a
$6\times 8$, and $\vec D$ a $6\times 6$ matrix. 
Matrix $\vec A$ has the form
\bea
\vec A = {\vec A}^{(0)} + {\vec A}^{(1)} \; ,
\eea
where ${\vec A}^{(0)}$ contains only bosonic fields and ${\vec A}^{(1)}$
is proportional to $\rho$. 
The explicit form of the matrices above is given 
in Appendix \ref{ListOfConstraints}. In order to see whether the matrix 
is singular, we have to consider the superdeterminant 
\cite{gitman90,cornwell89V3} 
\bea
{\rm sdet} \left\{ \phi^{(a)} , \phi^{(b)} \right\} & = &
{\rm det} \vec A \left[ {\rm det} 
\left(\vec D - \vec C {\vec A}^{-1} \vec B \right) \right]^{-1} \; .
\eea
Here we have
\bea
{\rm det} \vec A = -\left(1+2 \rho \right) \; ,
\eea
and
\bea
{\rm det} 
\left(\vec D - \vec C {\vec A}^{-1} \vec B \right) = 1 \; ,
\eea
such that the matrix of constraints is not singular. Hence we have only
second class constraints. 

Since $\left\{\phi^{(a)},\phi^{(b)} \right\}$ is not singular, it is possible 
to obtain the Lagrange multipliers from eq.\ (\ref{MotionOfConstraints})
by considering its inverse
\bea
\lambda_a & = & - \left\{H,\phi^{(b)}\right\} 
\left\{\phi^{(b)},\phi^{(a)} \right\}^{-1}
\; .
\eea 
Then, the equation of motion for an observable can be written in terms
of the Dirac bracket \cite{weinberg05V1}
\bea
\dot f = \left\{H,f\right\}_D \; ,
\eea
where 
\bea
\left\{f,g\right\}_D = \left\{f,g\right\} - \left\{f, \phi^{(a)}\right\}
\left\{\phi^{(a)},\phi^{(b)} \right\}^{-1} 
\left\{\phi^{(b)},g \right\} \; .
\eea

In order to obtain the inverse of the supermatrix of the constraints 
we determine first the inverses of ${\vec A}^{(0)}$ and $\vec D$, and
form the following matrix 
\bea
\vec N & = &
\left[
\begin{array}{cc}
\left({\vec A}^{(0)}\right)^{-1} & 0 \\
0 & {\vec D}^{-1}
\end{array}
\right] \; ,
\eea
Multiplying the matrix of the constraints (\ref{MatrixOfConstraints}) by
$\vec N$, we have
\bea
\left\{\phi^{(a)},\phi^{(b)}\right\} {\vec N} & = & \vec 1 - \vec R \, ,
\eea
where
\bea
\vec R & = & -
\left[
\begin{array}{cc}
{\vec A}^{(1)} 
\left({\vec A}^{(0)}\right)^{-1} & \vec B {\vec D}^{-1}\\
\vec C \left({\vec A}^{(0)}\right)^{-1} & 0
\end{array}
\right] \; .
\eea 
Then, since $\vec R$ contains Grassmann fields, the inverse of the matrix of 
constraints is achieved with a finite number of powers of $\vec R$:
\bea
\label{InvOfPoissonBrakets}
\left\{\phi^{(a)},\phi^{(b)}\right\}^{-1} & = &
{\vec N} \left( \vec 1 + \vec R + {\vec R}^2 + {\vec R}^3 + {\vec R}^4 \right)
\; . 
\eea

Having the inverse of the matrix of constraints, it can be readily seen that 
the commutations relations (\ref{Xalgebra}) are reproduced by the Dirac 
brackets: $\left[ X_i^{\alpha\beta} , X_j^{\gamma\delta} \right]_\pm 
= i \left\{ X_i^{\alpha\beta} , X_j^{\gamma\delta} \right\}_D$. 
Hence, the Lagrangian (\ref{Lagrangean}) and the set of constraints
(\ref{Constraint(1)}) lead to the algebra (\ref{Xalgebra}). 

The path integral quantization can be performed starting with the Hamiltonian,
and integrating over the canonical fields and the corresponding momenta.
As shown by Senjanovic \cite{senjanovic76,gitman90}, in the case of second 
class constraints, the path integral is as follows,
\bea
Z & = & \int {\cal D}X \, {\cal D} \Pi 
\prod_a \delta \left[ \phi^{(a)} \right] \,
{\rm sdet}^{1/2} \left\{ \phi^{(a)} , \phi^{(b)} \right\} \;
{\rm e}^{
i \int {\rm d} t \left[ \Pi \dot X - H \right] } \; .
\eea
In the present case it is possible to integrate over the momenta, leading to
\bea
Z & = & \int {\cal D} X \, \prod_i
\delta\left[ \phi^{(1)}_i \right] \, \delta\left[ \phi^{(2)}_i \right] \, 
\delta\left[ \phi^{(7)}_i \right] \, \delta\left[ \phi^{(9)}_i \right] \, 
\delta\left[ \phi^{(10)}_i \right] \, 
\left(1+2 \rho_i \right)^{\frac{1}{2}} \,
{\rm e}^{-S} \, ,
\eea
where the action in imaginary time is given by
\bea
S & = & \int_0^\beta {\rm d}\tau 
\Bigg\{
- \sum_i 
\frac{\left(1+\rho_i \right) u_i -1}
{\left(2-v_i\right)^2 - 4 \rho_i -u_i^2}
\left(X_i^{-+} {\dot X}_i^{+-} - X_i^{+-} {\dot X}_i^{-+} \right)
\nonumber \\ & & \qquad \qquad
+ \frac{1}{2} \sum_{i, \sigma} 
\left( X_i^{\sigma 0} {\dot X}_i^{0 \sigma} 
+  X_i^{0 \sigma} {\dot X}_i^{\sigma 0} \right) + H\left(X\right)
\Bigg\}
\; ,
\eea
with $\beta=1/k_B T$ the inverse temperature.
The term coming from the superdeterminant gives just a shift of the chemical
potential, and can be ignored. Since $X^{00}$ does not enter in the action,
we can integrate over it, eliminating the constraint $\phi^{(7)}$.

As a final step, we perform the change of variables introduced in 
\cite{wiegmann88,foussats99} that will be useful for further considerations:
\bea
\label{FromXtoOmegaPsi}
\begin{array}{rclrcl}
X^{++} & = & \left(1-\rho\right) \left(1+\Omega_z\right)/2 \; ,
&
X^{--} & = & \left(1-\rho\right) \left(1-\Omega_z\right)/2 \; ,
\\
X^{+-} & = & \left(1-\rho\right) \left(\Omega_x - i \Omega_y\right)/2 \; ,
&
X^{-+} & = & \left(1-\rho\right) \left(\Omega_x + i \Omega_y\right)/2 \; ,
\\
X^{+0} & = & \psi_+ \; ,
&
X^{-0} & = & \psi_- \; ,
\\
X^{0+} & = & \psi^*_+ \; ,
&
X^{0-} & = & \psi^*_- \; ,
\end{array}
\eea
where we introduced Grassmann field $\psi^{}_\pm$ and $\psi^*_\pm$. 
Accordingly, we have
$\rho=\psi^*_+ \psi^{}_+ + \psi^*_- \psi^{}_-$.
Since we have only 7 new variables, chosen in such a way that $\phi^{(1)}$
is automatically satisfied, we integrate over $X^{--}$ taking care of 
the constraint $\phi^{(1)}$ before performing the change of variables. After
the change of variables, 
the remaining constraints look as follows:
\bea
\label{OmegaConstraint}
\phi^{(2)} & = & \frac{1}{4} \left(1-\rho\right)^2 
\left( {\vec \Omega}^2 -1 \right) 
\; ,
\\
\label{ConstFermion1}
\phi^{(9)} & = & \psi^*_+ 
\frac{ \Omega_x - i \Omega_y}{1+\Omega_z} - \psi^*_- \; ,
\\
\label{ConstFermion2}
\phi^{(10)} & = & \psi_+ 
\left(\Omega_x + i \Omega_y\right) - \psi_- \left(1+\Omega_z \right) \; .
\eea 
For $\phi^{(2)}$ the factor $\left(1-\rho\right)^2$ can be 
absorbed in the chemical potential, such that it reduces to
$\phi^{(2)} \rightarrow {\tilde \phi}^{(2)} = 
\left( {\vec \Omega}^2 -1 \right)$.

After making the change of variables, we finally have
\bea
S & = & \int_0^\beta \Bigg\{ -\frac{i}{2} \sum_i 
\frac{\Omega_x {\dot \Omega}_y - \Omega_y {\dot \Omega}_x}{1+\Omega_z}
+ \sum_{i,\sigma} \psi^*_{i\sigma} {\dot \psi}^{}_{i\sigma}
+ t \sum_{<i,j> \atop \sigma} \psi^*_{i\sigma} \psi^{}_{j\sigma}
\nonumber \\ & & \qquad \quad
+ \frac{J}{8} \sum_{<i,j>} \left(1-\rho_i\right) \left(1-\rho_j\right)
\left( {\vec \Omega}_i \cdot {\vec \Omega}_j -1 \right)
- \mu \sum_i \rho_i 
\Bigg\}
\; ,
\eea
with constraints ${\tilde \phi}^{(2)}$, (\ref{ConstFermion1}), 
and (\ref{ConstFermion2}). For the undoped case, the action above reduces to 
the one corresponding to a quantum Heisenberg antiferromagnet, as obtained
e.g.\ using coherent states \cite{fradkin91}. 

\section{Staggered CP$^1$ representation
\label{StaggCP1}}
Here we concentrate us on the limit of low doping, in order to study the 
consequences of doping on the antiferromagnetic state present in the undoped
case.
In such a situation, we can assume a large correlation length for spins, such
that a long-wavelength expansion is justified. In this limit we can also
neglect terms $\sim \rho_i \rho_j$ obtaining thus a bilinear form in
the fermionic degrees of freedom. Then, the action has the following form
\bea
S & = & S_S + S_F \; ,
\eea
where
\bea
\label{Sspin}
S_S & = & \int_0^\beta {\rm d} \tau \Bigg\{ -\frac{i}{2} \sum_i 
\vec A [{\vec \Omega}_i] \cdot \partial_\tau {\vec \Omega}_i 
+ \frac{J}{8}
\sum_{<i,j>} {\vec \Omega}_i \cdot {\vec \Omega}_j 
\Bigg\}
\; ,
\eea
is the action of a pure Heisenberg model. $\vec A$ is the vector potential of 
a magnetic monopole, 
$\left(\vec \nabla \times \vec A \right) \cdot \vec \Omega =  1$, where 
derivatives are taken in the $\Omega$-space.
In this case it is given by
\bea
\label{GaugeBerry}
\vec A = \left[ - \frac{\Omega_y}{(1+\Omega_z)} , 
\frac{\Omega_x}{(1+\Omega_z)}, 0 \right]
\; .
\eea
The fermionic part is given by
\bea
\label{SF}
S_F & = & \int_0^\beta {\rm d} \tau \Bigg\{ 
\sum_{i,\sigma} \psi^*_{i\sigma} \partial_\tau \psi^{}_{i\sigma}
+ t \sum\limits_{<i,j> \atop \sigma} \psi^*_{i\sigma} \psi^{}_{j\sigma}
+ t' \sum\limits_{<<i,j>> \atop \sigma} 
\psi^*_{i\sigma} \psi^{}_{j\sigma}
\nonumber \\ & & \qquad \qquad
+ t'' \sum\limits_{<<<i,j>>> \atop \sigma} \psi^*_{i\sigma} \psi^{}_{j\sigma}
- \frac{J}{4} \sum_{<i,j> \atop \sigma}  \, \psi^*_{i\sigma} \psi^{}_{i\sigma}
{\vec \Omega}_i \cdot {\vec \Omega}_j
+ \mu \sum_{i,\sigma} \psi^*_{i\sigma} \psi^{}_{i\sigma}
\Bigg\} \; ,
\eea
where we allowed hopping to 2nd.\ nearest neighbors along the diagonals of the 
square lattice, denoted by $<<i,j>>$, with hopping amplitude $t'$, and 
to 2nd.\ nearest neighbors along the principal axis, denoted by
$<<<i,j>>>$, with hopping amplitude $t''$. Such hopping terms have been 
taken into account, since there is consensus that they are present in the
real materials \cite{kim98}. Furthermore, 
we have to take into
account the constraints ${\tilde \phi}^{(2)}$, (\ref{ConstFermion1})
and (\ref{ConstFermion2}). 

\subsection{Rotating reference frame and staggered CP$^1$ \\ representation}
In order to take into account the constraints (\ref{ConstFermion1})
and (\ref{ConstFermion2}), and in
order to work with smoothly varying fields, we consider now the fact that
the field $\vec \Omega$ is staggered on nearest neighbors and define
a local quantization axis for the fermions with a rotation that 
fulfills
\bea
\label{spinrot'}
U^\dagger_i \, {\vec \Omega}^{}_i \cdot \vec \sigma \, U^{}_i = 
(-1)^i \, \sigma^z \; ,
\eea 
where $\sigma^a$, with $a=x$, $y$, or $z$ are the Pauli matrices.
For $i$ even the condition above is accomplished by $U \in$ SU(2),
\bea
\label{matrixU}
U =
\left(
\begin{array}{rr}
z_1 & -z_2^* \\
z_2 & z_1^*
\end{array}
\right) \; ,
\eea
where $\bar z \, z = 1$, and
\bea
\label{Omega-cp1}
\Omega^a = \bar z \, \sigma^a \, z \; .
\eea
For $i$ odd, on the other hand, we take rotation matrices (\ref{matrixU}) and
\bea
\label{cp1stag}
\Omega^a = z^{}_\alpha \, \sigma^y_{\alpha \beta} \, 
\sigma^a_{\beta \gamma} \, \sigma^y_{\gamma \delta} \, z^*_\delta \; .
\eea 
This ensures that (\ref{spinrot'}) is fulfilled.

\subsubsection{Constraints in the rotating reference frame\label{RotatedFermions}}
Here we discuss the transformation of the constraints (\ref{ConstFermion1})
and (\ref{ConstFermion2}) on going to the rotating reference frame introduced
above. For simplicity of notation,we redefine them as follows.
\bea
\label{ConstraintsBeforeRotation}
\varphi^*_{F,2} & = & -\psi^*_+ 
\frac{ \Omega_x - i \Omega_y}{1+\Omega_z} + \psi^*_- \; ,
\nonumber \\
\varphi_{F,2} & = & -\psi_+ 
\left(\Omega_x + i \Omega_y\right) + \psi_- \left(1+\Omega_z \right) \; ,
\eea 
Since $\varphi^*_{F,2}$ is a fermionic constraint, we can use a corresponding 
$\delta$-function for Grassmann variables $\xi$ and $\xi'$:
\bea
\delta \left( \xi - \xi' \right) = -\left( \xi - \xi' \right) \; ,
\eea
such that 
\bea
\delta \left( \varphi^*_{F,2} \right) = \frac{1}{1+\Omega_z} \,
\delta \left( {\tilde \varphi}^*_{F,2} \right) \; ,
\eea
where we defined ${\tilde \varphi}^*_{F,2} \equiv -\psi^*_+ 
\left(\Omega_x - i \Omega_y\right)
+ \psi^*_- \left(1+\Omega_z\right)$.
Furthermore,
going back to the original formulation in terms of the $X$-fields, we can
rewrite $\phi^{(9)}$ in (\ref{Constraint(1)}) as follows. Insertion of 
$\phi^{(1)}$ into $\phi^{(2)}$ leads to
\bea
\phi^{(2')} & = & X^{+-} X^{-+} - X^{++} X^{--}
\; ,
\eea
such that
\bea
X^{++} & = & \frac{X^{+-} X^{-+}}{X^{--}} \; ,
\eea
and $\phi^{(9)}$ can be brought to the form
\bea
\label{phi9'}
\phi^{(9')} & = & X^{0+} X^{--} - X^{0-} X^{-+} \; ,
\eea
that 
after the change of variables (\ref{FromXtoOmegaPsi}) we can write as
\bea
{\tilde \varphi}^*_{F,1} & = &  \psi^*_+ 
\left(1-\Omega_z\right) - \psi^*_- \left(\Omega_x + i \Omega_y \right)\; .
\eea
Then, we can consider the first constraint in (\ref{ConstraintsBeforeRotation})
as part of
\bea
{\tilde \varphi}^*_F = 
\psi^* \left( \vec 1 - \vec \Omega \cdot \vec \sigma \right) = 0
\; .
\eea
We can introduce now new fermions $\chi = U^\dagger \psi$ and 
$\chi^*=\psi^*U$, such that 
\bea
{\tilde \varphi}^*_F U = \chi^* U^\dagger 
\left( \vec 1 - \vec \Omega \cdot \vec \sigma \right) U 
= \chi^* \left(\vec 1 \mp \sigma^z \right) = 0 \; ,
\eea
where the upper (lower) sign is for even (odd) sites. 

In the same way, we have for the second constraint in 
(\ref{ConstraintsBeforeRotation})
\bea
U^\dagger \varphi_F = U^\dagger 
\left( \vec 1 - \vec \Omega \cdot \vec \sigma \right) U \chi
= \left(\vec 1 \mp \sigma^z \right) \chi = 0 \; .
\eea

However, since the constraints are actually given solely in terms of 
${\tilde \varphi}^*_F$ and $\varphi^{}_F$, we introduce 
\bea
\theta^* \equiv {\tilde \varphi}^*_F U \; , \quad
\theta \equiv U^\dagger \varphi_F \; ,
\eea
such that for the original form of the constraints we have
\bea
\delta \left( \varphi^*_{F,2} \right) \, 
\delta \Big( \varphi^{}_{F,2} \Big) & = &
\frac{1}{1+\Omega_z} \,
\delta \left( {\tilde \varphi}^*_{F,2} \right) \,
\delta \left( \varphi^{}_{F,2} \right)
\nonumber \\ & = &
\frac{1}{1+\Omega_z} \,
\delta \left( \theta^*_1 z^*_2 + \theta^*_2 z^{}_1 \right) \,
\delta \left( z_2 \theta^{}_1 + z^*_1 \theta^{}_2 \right) \; .
\eea

We consider now the action of the constraints on even and on odd sites.
\begin{itemize}

\item[{\it i})] Even sites.
\bea
\theta^*_1 = \theta^{}_1 = 0 \; , \quad 
\theta^*_2 = 2 \chi^*_- \; , \quad 
\theta^{}_2 = 2 \chi^{}_- \; .
\eea
Furthermore, we have
\bea
1+\Omega_z = 2 \mid z_1 \mid^2 \; ,
\eea
such that finally,
\bea
\delta \left( \varphi^*_{F,2} \right) \, 
\delta \Big( \varphi^{}_{F,2} \Big) = 
\frac{1}{2 \mid z_1 \mid^2} \, 
\delta \left( \theta^*_2 z^{}_1 \right) \,
\delta \left( z^*_1 \theta^{}_2 \right)
= 2 \delta \left( \chi^*_- \right) \, \delta \left( \chi^{}_- \right)
\, .
\eea

\item[{\it i})] Odd sites.
\bea
\theta^*_1 = 2 \chi^*_+ \; , \quad 
\theta^{}_1 = 2 \chi^{}_+  \; , \quad 
\theta^*_2 = \theta^{}_2 = 0 \; .
\eea
In this case we have due to (\ref{cp1stag}), 
\bea
1+\Omega_z = 2 \mid z_2 \mid^2 \; ,
\eea
such that finally,
\bea
\delta \left( \varphi^*_{F,2} \right) \, 
\delta \Big( \varphi^{}_{F,2} \Big) = 
\frac{1}{2 \mid z_2 \mid^2} \, 
\delta \left( \theta^*_1 z^*_2 \right) \,
\delta \left( z^{}_2 \theta^{}_1 \right)
= 2 \delta \left( \chi^*_+ \right) \, \delta \left( \chi^{}_+ \right)
\, .
\eea

\end{itemize}

The constraints above lead to $\chi_- =0$ on even sites, whereas 
$\chi_+ =0$ on odd sites.
We can therefore work with spinless fermions $\chi_A$ on even sites and
$\chi_B$ on odd sites, where $A$ and $B$ denote the two sublattices,
such that the constraints on fermions are exactly taken into account.

\subsubsection{Slow CP$^1$ variables in the rotating reference frame\label{SlowCP1}} 
After introducing cells $j$, each one containing one even ($A$) and one odd 
($B$) site,  we
can define new fields
\bea
\label{newz}
{\tilde z}_j = \frac{1}{2} \left( z_j^B + z_j^A \right) \; , \quad
a \zeta_j = \frac{1}{2} \left( z_j^B - z_j^A \right) \; ,
\eea
where $a$ is the original lattice constant. Due to the constraints 
$\mid z^A_j \mid^2 = \mid z^B_j \mid^2 = 1$, the new fields are subjected to
the constraint 
\bea
\label{constzz}
\bar {\tilde z} \, \zeta + \bar \zeta \, \tilde z = 0 \; .
\eea
This condition can be used to fix the phase of $\zeta$ with respect to
that of $z$, such that they change by the same amount under a gauge
transformation. This will be discussed in more detail in Sec.\ 
\ref{GaugeFixing}.
Equation (\ref{constzz}) also implies that both fields
are subjected to the constraint
\bea
\label{modzz}
\bar {\tilde z} \, \tilde z + a^2 \, \bar \zeta \, \zeta = 1 \; .
\eea
We then introduce
new fields 
\bea
\label{newnewz}
\tilde z = z \sqrt{1 - a^2 \, \bar \zeta \zeta} \; ,
\eea 
such that the constraint (\ref{modzz}) is satisfied with $\bar z z = 1$,
and the constraint (\ref{constzz}) translates into
\bea
\label{ConstZZFinal}
\bar z \, \zeta + \bar \zeta \, z = 0 \; .
\eea

The fact that $z_j^B - z_j^A$ is of ${\cal O}(a)$ can also be seen by 
going back to the vector representation, where we have 
${\vec \Omega}^A - {\vec \Omega}^B \sim \bar z \vec \sigma z$
while ${\vec \Omega}^A + {\vec \Omega}^B \sim \bar \zeta \vec \sigma z
+ \bar z \vec \sigma \zeta$, i.e.\ the field $\zeta$ is directly related 
to ferromagnetic fluctuations within the unit cell.

\section{Long-wavelength expansion}
Once we have identified smoothly varying fields and their slow modes, we 
perform an expansion of the action in powers of the lattice constant $a$
up to second order, after a transformation to the rotating reference frame.
For clarity of the presentation
we deal first with the action $S_S$ in eq.\ (\ref{Sspin}) and then with
$S_F$ in eq.\ (\ref{SF}).

\subsection{\label{spinaction}
Spin action in the staggered CP$^1$ representation}
Here we consider the pure Heisenberg model as given by (\ref{Sspin}). 
We pass to CP$^1$ variables using (\ref{Omega-cp1}) for even sites and
(\ref{cp1stag}) for odd sites. 

For the Berry phase we have
\bea
\label{BerryAF'}
-\sum_i \frac{i}{2} 
\vec A [{\vec \Omega}_i] \cdot \partial_\tau {\vec \Omega}_i 
= 2 a \sum_j \left[ {\bar z}_j \, \partial_\tau \zeta_j 
+ {\bar \zeta}_j \, \partial_\tau z_j \right] 
+ {\cal O} \left(a^4 \right)
\; .
\eea
Terms containing a total time derivative were discarded due to periodic 
boundary conditions in imaginary time.

For the interaction term we first discuss our convention in defining new 
coordinates for the units cells containing sites $A$ and $B$. On passing to 
the new coordinate system we choose
\bea
x' = \frac{1}{\sqrt{2}} \left( x + y \right) \; , \qquad
y' = \frac{1}{\sqrt{2}} \left( y - x \right) \; ,
\eea
such that in the new coordinate system the basis vectors for sublattice A and 
B are, respectively,
\bea
\label{BasisVectors}
{\vec x}_A = (0,0) \; , \quad
{\vec x}_B = \left(a/\sqrt{2},-a/\sqrt{2}\right) \; .
\eea

Introducing the notation
\bea
\label{DefGAndF}
G_j \equiv 2 i z_{j\alpha} \sigma^y_{\alpha \beta} \zeta_{j\beta} \; , \qquad
F_{j\mu} \equiv  i z_{j\alpha} \sigma^y_{\alpha \beta} \partial_\mu z_{j\beta}
\; ,
\eea 
where $\mu = x$ or $y$, we obtain after a lengthy but straightforward 
calculation,
\bea
\label{SumOmegaiOmegaj}
\sum_{<i,j>} {\vec \Omega}_i \cdot {\vec \Omega}_j & = &
8 \int {\rm d}x^2 
\Bigg\{
2\left[ G_j G^*_j + F_{jy} F^*_{jy} + F^{}_{jx}  F^*_{jx} \right]
- \left( F^{}_{jx} F^*_{jy} + F^{}_{jy} F^*_{jx} \right)
\nonumber \\ & & \qquad \qquad \quad
+ \sqrt{2} \left[ \left( F^{}_{jy} -  F^{}_{jx} \right) G^*_j 
+ G^{}_j \left( F^*_{jy} - F^*_{jx} \right) \right]
\Bigg\}
\; ,
\eea
where constants terms were discarded.

\subsection{Fermionic part in the staggered CP$^1$ representation \label{FermionicPart}}
Here we consider the action (\ref{SF}) in the rotating reference frame, i.e.\
with fermions as defined in Sec.\ \ref{RotatedFermions}:
\bea
\chi = U^\dagger \psi \; , \quad \chi^*=\psi^*U \; .
\eea
After applying the constraints (\ref{ConstFermion1}) and (\ref{ConstFermion2}),
we can define a new spinor per unit cell
\bea
\label{spinor-chi}
\chi =
\left(
\begin{array}{c}
\chi_A \\ \chi_B
\end{array}
\right) \; ,
\eea 
following the discussion in Sec.\ \ref{RotatedFermions}.

\subsubsection{Temporal derivatives} 
Defining 
\bea
A_{j,\mu} = -i {\bar z}_j \partial_\mu z_j \; , \qquad
C_{j,\mu} \equiv {\bar z}_j \, \partial_\mu \zeta_j
+ {\bar \zeta}_j \, \partial_\mu z_j \; ,
\eea
we have
\bea
\sum_{i,\sigma} \psi^*_{i\sigma} \partial_\tau \psi^{}_{i\sigma} & = &
\sum_j \chi^*_j 
\left( D_\tau - a C_\tau \right) \chi^{}_j 
\; ,
\eea
where we introduced
\bea
\label{CovDerivA}
D_\mu & \equiv & \partial_\mu + i \sigma^z A_\mu
\; .
\eea
At this point we can check gauge invariance. Since $\chi_A$ and $\chi_B$
have opposite charges, a gauge transformation leads to
\bea
\label{gaugetr}
\begin{array}{rclcrcl}
\chi_A & \rightarrow & 
{\rm e}^{+i \phi} \, \chi_A \; , & & 
\chi_B & \rightarrow & 
{\rm e}^{-i \phi} \, \chi_B \; ,
\\
z & \rightarrow & 
{\rm e}^{+i \phi} \, z \; , & & 
\zeta & \rightarrow & 
{\rm e}^{+i \phi} \, \zeta \; .
\end{array}
\eea
It can be easily seen that the term with the covariant derivative
is invariant under the gauge transformations above, and
that $C_\mu$ is gauge invariant due to the constraint (\ref{constzz}):
\bea
C_\mu & \rightarrow & 
{\bar z} \, \partial_\mu \zeta
+ {\bar \zeta} \, \partial_\mu z
+ \partial_\mu \phi \left( \bar z \, \zeta + \bar \zeta \, z \right) \; .
\eea

\subsubsection{Kinetic term}
For the kinetic term of the fermions we have the contributions proportional to
$t$, $t'$, and $t''$, that we treat separately in the following. We consider 
first contributions from nearest neighbor hopping, where, after Fourier
transformation, we have
\bea
\label{KineticTermFinal}
- t \sum\limits_{<i,k> \atop \sigma} 
\psi^*_{i\sigma} \psi^{}_{k\sigma} & \rightarrow &
-t \sum_{k_1,k_2} \chi^*_{k_1} \, \Xi ({\vec k}_1, {\vec k}_2 ) \, \chi^{}_{k_2} \; ,
\eea
where
\bea
\label{XiAll}
\Xi ({\vec k}_1, {\vec k}_2 ) & \equiv &
\left[ \sum_{i=1}^4 \Xi^{(i)}_{AB} ({\vec k}_1, {\vec k}_2 ) \right] \sigma^+ 
+\left[ \sum_{i=1}^4 \Xi^{(i)}_{BA} ({\vec k}_1, {\vec k}_2 ) \right] \sigma^- 
\; ,
\eea
with $\sigma^\pm = \left( \sigma^x \pm i \sigma^y \right)/2$.
The explicit expressions for $\Xi^{(i)} ({\vec k}_1, {\vec k}_2 )$ are given 
in Appendix \ref{HoppingCP1}. They contain only 
contributions up to ${\cal O} (a)$, since, as seen above, terms 
proportional to $t$ are off-diagonal, and hence, no contribution linear
in $\Xi$ appears in the final action, as shown in Sec.\ 
\ref{IntegrateOutFermions}. They enter in a quadratic form after integrating
fermions out. 

Next we consider contributions from second nearest neighbor hopping along the 
diagonal. In this case we have
\bea
\label{KineticTPrimeFinal}
- t' \sum\limits_{<<i,k>> \atop \sigma} 
\psi^*_{i\sigma} \psi^{}_{k\sigma} & \rightarrow &
\sum_{k_1,k_2} \chi^*_{k_1} \, 
\Psi ({\vec k}_1, {\vec k}_2 ) \, \chi^{}_{k_2} \; ,
\eea
where 
\bea
\label{TildePsiAll}
\Psi ({\vec k}, {\vec k}' ) & = &
\epsilon_1 \left({\vec k}\right) \, \delta_{\vec k , {\vec k}'} \, \vec 1 +
\left[
\sum_{i=1}^4 \Psi^{(i)}_{AA} ({\vec k}, {\vec k}' ) \right] \gamma^+ 
+ \left[ 
\sum_{i=1}^4 \Psi^{(i)}_{BB} ({\vec k}, {\vec k}' ) \right] \gamma^- 
\; ,
\eea 
with
\bea
\label{epsilon1}
\epsilon_1 \left({\vec k}\right) & = & 
-2 t' \left[ \cos \left(\sqrt{2} k_x a \right) 
+ \cos \left(\sqrt{2} k_y a \right) \right] \; ,
\eea 
and $\gamma^\pm = \left(\vec 1 \pm \sigma^z \right)/2$. Again, explicit 
expressions for $\Psi^{(i)} ({\vec k}, {\vec k}' )$ are given
in Appendix \ref{HoppingCP1}. Contrary to the previous case, we consider 
here terms up to 
${\cal O}\left(a^2\right)$, since terms proportional to $t'$ connect sites 
within a sublattice,
giving rise to diagonal contributions to the self-energy of the fermions, as 
seen above.

Finally, we have the contributions from second nearest neighbor hopping along 
the principal axis:
\bea
\label{KineticTDoublePrimeFinal}
- t'' \sum\limits_{<<<i,k>>> \atop \sigma} 
\psi^*_{i\sigma} \psi^{}_{k\sigma} & \rightarrow &
\sum_{k_1,k_2} \chi^*_{k_1} \, 
\Phi ({\vec k}_1, {\vec k}_2 ) \, \chi^{}_{k_2} \; ,
\eea
where 
\bea
\label{TildePhiAll}
\Phi ({\vec k}, {\vec k}' ) & = &
\epsilon_2 \left({\vec k}\right) \, \delta_{\vec k , {\vec k}'} \, \vec 1 +
\left[
\sum_{i=1}^4 \Phi^{(i)}_{AA} ({\vec k}, {\vec k}' ) \right] \gamma^+ 
+ \left[ 
\sum_{i=1}^4 \Phi^{(i)}_{BB} ({\vec k}, {\vec k}' ) \right] \gamma^- 
\; ,
\eea 
with
\bea
\label{epsilon2}
\epsilon_2 \left({\vec k}\right) & = & 
-4 t'' \cos \left(\sqrt{2} k_x a \right) 
\cos \left(\sqrt{2} k_y a \right)  \; ,
\eea 
and $\Phi^{(i)} ({\vec k}, {\vec k}' )$ as given in Appendix 
\ref{HoppingCP1}. Here again, contributions up to ${\cal O} \left(a^2\right)$
have to be taken into account.

\subsubsection{Spin interaction dressed with fermions}
For such contributions we have
\bea
\frac{J}{4} \sum_{<i,k> \atop \sigma} \, 
\psi^*_{i\sigma} \psi^{}_{i\sigma} \, {\vec \Omega}_i \cdot {\vec \Omega}_k
& = &
\frac{J}{4} \sum_{k_1,k_2} \chi^*_{k_1} \, {\cal F} ({\vec k}_1, {\vec k}_2 ) 
\, \chi^{}_{k_2} 
- J \sum_k \chi^*_k \, \chi^{}_k 
\; ,
\eea
where 
\bea
\label{CalF}
{\cal F} ({\vec k}_1, {\vec k}_2 ) & \equiv &
{\cal F}^+   \, \vec 1 +  {\cal F}^-\, \sigma^z  \; ,
\eea
with ${\cal F}^+$ and ${\cal F}^-$ displayed in Appendix \ref{HoppingCP1}.

\section{Integration of fermionic degrees of freedom\label{IntegrateOutFermions}}
Once the gradient expansion was performed, we integrate out in a first step
the fermionic degrees of freedom, that are considered at all wavelengths.
In a second step we integrate out the magnetic fast degrees 
of freedom in order to obtain the effective action for the slow magnetic 
modes. 

Collecting all results obtained in Sec.\ \ref{FermionicPart} we have the 
following form for the action
\bea
S_F & = & 
-\sum_{k,k'} \chi^*_k
\left[  G_0^{-1} (k,k') - \Sigma (k,k') \right] \chi^{}_{k'}
\; ,
\eea
where $k = (i\nu_n, \vec k )$, $\nu_n$ being Matsubara frequencies for 
fermions, and we defined 
\bea
G_0^{-1} (k,k') \equiv \frac{1}{\beta} \left\{ i \nu_n 
- \left[\epsilon \left(\vec k \right)+J+\mu\right] \right\} 
\delta_{kk'} \; ,
\eea
i.e.\ the free propagator for fermions in the present theory.
The self-energy $\Sigma$ is given by
\bea
\label{SigmaAll}
\Sigma (k,k') & = & 
i \sigma^z A_\tau (k, k') - a C_\tau (k,k')
+ t \, \Xi (k,k') 
+ t' \, \Psi (k,k') + t'' \, \Phi (k,k')
- \frac{J}{4} {\cal F} (k,k') \; .
\eea

The dispersion relation for 
the holes is given by the ones in (\ref{epsilon1}) and (\ref{epsilon2}).
\bea
\epsilon \left(\vec k \right) = \epsilon_1 \left(\vec k \right) +
\epsilon_2 \left(\vec k \right) 
= -2 t' \left[ \cos \left(\sqrt{2} k_x a \right) 
+ \cos \left(\sqrt{2} k_y a \right) \right] 
-4 t'' \cos \left(\sqrt{2} k_x a \right) \, 
\cos \left(\sqrt{2} k_y a \right) \; .
\eea
This dispersion is obtained for the magnetic Brillouin zone. By rotating to 
the original Brillouin zone, we have
\bea
k_x' = \frac{1}{\sqrt{2}} \left(k_x - k_y \right) \; , \quad
k_y' = \frac{1}{\sqrt{2}} \left(k_x + k_y \right) \; ,
\eea
and the dispersion is given by
\bea
\label{epsilonkPrime}
\epsilon \left({\vec k}' \right) & = & 
-4 t' \cos \left(k_x a \right) \, \cos \left(k_y a \right) 
-2 t'' \left[ \cos \left(2 k_x a \right) 
+ \cos \left(2 k_y a \right) \right] \; .
\eea
It should be noticed, that the free dispersion of the dopant holes is 
determined here entirely by $t'$ and $t''$. This is to be contrasted
with calculations of the single hole dispersion of the pure $t$-$J$ model
in the self-consistent Born approximation \cite{kane89,martinez91,liu92}
that are in very good agreement with quantum Monte Carlo simulations 
\cite{brunner00b}, where a dispersive band is found also in the case
$t'=t''=0$. Instead, in our case, on passing to the continuum limit, 
such a dispersion can be obtained only by explicitely introducing 
finite values for those hopping amplitudes. In fact, 
the dispersion of the single hole found previously can
be reproduced by (\ref{epsilonkPrime}) for appropriate values of $t'$ and 
$t''$ ($t'=-0.3t$ and $t''=-0.115t$ for $J=0.4t$ in Ref.\ \cite{brunner00b}). 
Hence, on focusing on lower energy scales, operators corresponding to
$t'$ and $t''$ should be expected to be generated even if such operators are
initially missing. 

Arranging terms according to powers of the UV cutoff $a$, we have
\bea
\Sigma(k,k') & = & a \Sigma^{(1)} (k,k') + a^2 \Sigma^{(2)} (k,k')
+ {\cal O} \left(a^3\right) \; ,
\eea 
such that after integrating out fermions and keeping contributions up to 
${\cal O} \left(a^2\right)$, the fermionic part of the action goes over into
\bea
S_F & \rightarrow & 
- {\rm Tr} \left( G_0 \Sigma + \frac{1}{2} G_0 \Sigma G_0 \Sigma \right)
+ {\cal O} \left(a^3 \right) 
\; .
\eea
The terms entering (\ref{SigmaAll}) can be regrouped as
\bea
\Sigma^{(i)} & = &  \Sigma^{(i)}_0 \, \vec 1 + \Sigma^{(i)}_z
\, \sigma^z + \Sigma^{(i)}_+ \, \sigma^+ +
\Sigma^{(i)}_- \, \sigma^- 
\; , 
\eea
with $i=1,2$. This leads for the terms linear in $\Sigma$ to
\bea
\label{ExpansionInG0Sigma}
{\rm Tr} \, G_0 \Sigma & = &  {\rm Tr} \, G_0 
\left[a \Sigma^{(1)}_0  + a^2 \Sigma^{(2)}_0 \right] 
+ {\cal O} \left(a^3 \right)\; ,
\eea
and for the terms quadratic in $\Sigma$ to
\bea
\label{G0SigmaG0Sigma}
{\rm Tr} \,  G_0 \Sigma G_0 \Sigma & = & 
a^2 {\rm Tr} \, 
\left[G_0 \Sigma^{(1)}_0 G_0 \Sigma^{(1)}_0
+ G_0 \Sigma^{(1)}_z G_0 \Sigma^{(1)}_z 
+ 2 G_0 \Sigma^{(1)}_+ G_0 \Sigma^{(1)}_-
\right] + {\cal O} \left(a^3 \right)\; .
\eea
Introducing indices $\tau$, $t$, $t'$, and $t''$ for the temporal, and
different hopping processes, respectively, we arrange the different 
contributions in (\ref{SigmaAll})
as follows:
\bea
\Sigma^{(i)}_{0,z} (k,k') & = & 
\Sigma^{(i,\tau)}_{0,z} + \Sigma^{(i,t')}_{0,z} + \Sigma^{(i,t'')}_{0,z} \; ,
\eea
while $\Sigma^{(1)}_{\pm}$ contains only contributions proportional to $t$. 

An explicit evaluation using the results in Appendix \ref{HoppingCP1}
shows that there are no contributions in first order in $a$. 
The contributions in 
${\cal O} \left(a^2\right)$ coming from ${\rm Tr} \, G_0 \Sigma$ 
are as follows:
\bea
{\rm Tr} \, G_0 \Sigma^{(2,\tau)} & = & \frac{4{\tilde \rho}}{a} \int_0^\beta {\rm d} \tau 
\int {\rm d}^2 x \, \left( {\bar z} \partial_\tau \zeta 
+ {\bar \zeta} \partial_\tau z \right)
\; ,
\nonumber \\
{\rm Tr} \, G_0 \Sigma^{(2,t')} & = & 
-16t'' {\tilde \rho}_1 \int {\rm d}\tau \int {\rm d}^2 x \, 
\left(
\partial_x \bar z \partial_x z 
+ \partial_y \bar z \partial_y z 
\right)
\; ,
\nonumber \\
{\rm Tr} \, G_0 \Sigma^{(2,t'')} & = & 
-16t'' {\tilde \rho}_2 \int {\rm d}\tau \int {\rm d}^2 x \,  
\left(\partial_x \bar z \partial_x z 
+ \partial_y \bar z \partial_y z\right) \; ,
\nonumber \\
{\rm Tr} \, G_0 \Sigma^{(2,J)} & = & 
-8J\tilde \rho \int {\rm d}\tau \int {\rm d}^2 x \,
\Bigg\{
2 G G^* + 2 F^{}_{y} F^*_{y} 
+ 2 F^{}_{x} F^*_{x} 
- \left( F^{}_{y} F^*_{x} + F^{}_{x} F^*_{y} \right)  
\nonumber \\ & & \qquad \qquad \qquad \qquad \quad +
\sqrt {2} \Bigg[ \left( F^{}_{y} - F^{}_{x} \right) 
G^* + G  \left(F^*_{y} -  F^*_{x} \right) \Bigg]
\Bigg\}
\; .
\eea
where $\tilde \rho$ is the density of holes, i.e.\
\bea
\tilde \rho = \frac{1}{N} 
\sum_{\vec k} n\left(\vec k\right) \; ,
\eea
${\tilde \rho}_1$ is defined
as
\bea
{\tilde \rho}_1 \equiv \frac{1}{N} 
\sum_{\vec k} n\left(\vec k\right) \, 
\cos \left(\sqrt{2} k_x a\right)
\eea
and 
\bea
{\tilde \rho}_2 \equiv \frac{1}{N}
\sum_{\vec k} n\left(\vec k\right) \, 
\cos \left(\sqrt{2} k_{x} a \right) \, \cos \left(\sqrt{2} k_{y} a \right)
\; .
\eea
In the expressions above, and in the following, $n\left(\vec k \right)$
is the Fermi distribution function
\bea
n\left(\vec k \right) = \frac{1}
{\exp \left\{\beta\left[\epsilon \left(\vec k \right) - \mu \right]\right\}
+ 1} \; . 
\eea 

We focus now on the contributions quadratic in $\Sigma$, given in eq.\
(\ref{G0SigmaG0Sigma}).
The different terms from the expressions containing $\Sigma^{(1)}_0$ are:
\bea
{\rm Tr} \, G_0 \Sigma^{(1)}_0 G_0 \Sigma^{(1)}_0 & = &
{\rm Tr} \, G_0 \Sigma^{(1,\tau)}_0 G_0 \Sigma^{(1,\tau)}_0
+ 2 {\rm Tr} \, G_0 \Sigma^{(1,\tau)}_0 G_0 
\left[ \Sigma^{(1,t')}_0 + \Sigma^{(1,t'')}_0 \right]
\nonumber \\ & &
+ {\rm Tr} \, G_0 \Sigma^{(1,t')}_0 G_0 \Sigma^{(1,t')}_0
+ 2 {\rm Tr} \, G_0 \Sigma^{(1,t')}_0 G_0 \Sigma^{(1,t'')}_0 
+ {\rm Tr} \, G_0 \Sigma^{(1,t'')}_0 G_0 \Sigma^{(1,t'')}_0 \; .
\eea
The same expression results for the terms containing $\Sigma^{(1)}_z$.

In the following we consider the different nonvanishing contributions.
From the temporal part we have  
\bea
{\rm Tr} \, G_0 \Sigma^{(1,\tau)}_0 G_0 \Sigma^{(1,\tau)}_0
+{\rm Tr} \, G_0 \Sigma^{(1,\tau)}_z G_0 \Sigma^{(1,\tau)}_z & = &
\frac{4\tilde \rho \, \kappa}{a^2}  
\int_0^\beta {\rm d}\tau \int {\rm d}^2 x
\, \left({\bar z} \, \partial_\tau z \right)^2 \; ,
\eea
where the electronic compressibility is defined as usually,
\bea
\kappa & \equiv & \frac{1}{\tilde \rho} 
\frac{\partial \tilde \rho}{\partial \mu} \; .
\eea
From terms proportional to $t$ we have
\bea
{\rm Tr} \, G_0 \Sigma^{(1,t)}_+ G_0 \Sigma^{(1,t)}_- & = &
16 t^2 \tilde \rho 
\int_0^\beta {\rm d}\tau \int {\rm d}^2x 
\left[
2 \kappa_1  G^* G
+ \sqrt{2} \kappa_1 G^* \left( F_y -F_x \right) 
+ \sqrt{2} \kappa_1  
\left( F^*_y -F^*_x \right) G \right.
\nonumber \\ & & \qquad \qquad \qquad \qquad \quad
\left.
+ \left(\kappa_2 F^*_y F^{}_y - \kappa_1 F^*_y F^{}_x
- \kappa_1 F^*_x F^{}_y + \kappa_2 F^*_x F^{}_x 
\right)
\right] \; ,
\eea
where we defined
\bea
\label{DefsKappas}
\kappa_1 & \equiv & \frac{1}{\tilde \rho} \frac{\partial}{\partial \mu}
\frac{1}{N} \sum_{\vec k} n\left(\vec k \right) \,
\cos^2 \left(\frac{\sqrt{2}}{2} k_{x} a \right) \, 
\cos^2 \left(\frac{\sqrt{2}}{2} k_{y} a \right) \; ,
\eea
and
\bea
\kappa_2 & \equiv & \frac{1}{\tilde \rho} \frac{\partial}{\partial \mu}
\frac{1}{N} \sum_{\vec k} n\left(\vec k \right) \,
\cos^2 \left(\frac{\sqrt{2}}{2} k_{x} a \right) 
\; .
\eea
The terms proportional to $t'$ lead to
\bea 
{\rm Tr} \, G_0 \Sigma^{(1,t')}_0 G_0 \Sigma^{(1,t')}_0
+{\rm Tr} \, G_0 \Sigma^{(1,t')}_z G_0 \Sigma^{(1,t')}_z & = &
-32 {t'}^2 a^2 \tilde \rho \kappa_3 \sum_j 
\int_0^\beta {\rm d}\tau \, \left[
\left( \bar z \partial_x z \right)^2 
+ \left( \bar z \partial_y z \right)^2
\right] 
\; ,
\eea
where we defined
\bea
\label{DefKappa3}
\kappa_3 & \equiv & \frac{1}{\tilde \rho} \frac{\partial}{\partial \mu}
\frac{1}{N} \sum_{\vec k} n\left(\vec k \right) \,
\sin^2 \left(\sqrt{2} k_{x} a \right) 
\; .
\eea
From the terms proportional to $t''$ we obtain
\bea
{\rm Tr} \, G_0 \Sigma^{(1,t'')}_0 G_0 \Sigma^{(1,t'')}_0
+{\rm Tr} \, G_0 \Sigma^{(1,t'')}_z G_0 \Sigma^{(1,t'')}_z & = &
- 64 {t''}^2a^2 \tilde \rho \kappa_4 \sum_j
\int_0^\beta {\rm d}\tau \, \left[ 
\left( \bar z \partial_x z \right)^2
+ \left( \bar z \partial_y z \right)^2 \right]
\; ,
\eea
with
\bea
\label{DefKappa4}
\kappa_4 & \equiv & \frac{1}{\tilde \rho} \frac{\partial}{\partial \mu}
\frac{1}{N} \sum_{\vec k} n\left(\vec k \right) \,
\sin^2 \left(\sqrt{2} k_{x} a \right) \, 
\cos^2 \left(\sqrt{2} k_y a\right) \; .
\eea
Finally, there is a contribution proportional to $t' t''$ of the form
\bea
{\rm Tr} \, G_0 \Sigma^{(1,t')}_0 G_0 
\Sigma^{(1,t'')}_0 
+{\rm Tr} \, G_0 \Sigma^{(1,t')}_z G_0 \Sigma^{(1,t'')}_z & = &
-64 t' t'' a^2 \tilde \rho \kappa_5 \sum_j
\int_0^\beta {\rm d}\tau \left[ \left( \bar z \partial_x z \right)^2
+ \left( \bar z \partial_y z \right)^2 \right] \; ,
\eea
where
\bea
\label{DefKappa6}
\kappa_5 & \equiv & \frac{1}{\tilde \rho} \frac{\partial}{\partial \mu}
\frac{1}{N} \sum_{\vec k} n\left(\vec k \right) 
\bigg[
\sin^2 \left(\sqrt{2} k_{x} a \right) \, 
\cos \left(\sqrt{2} k_y a\right) 
+ \sin^2 \left(\sqrt{2} k_y a \right) \, 
\cos \left(\sqrt{2} k_x a\right) 
\bigg]
\; ,
\eea

Collecting all contributions after integrating fermions out, we have
\bea
S_F & \rightarrow & S_F^{(z)} + S_F^{(\zeta)}
\; ,
\eea
where 
\bea
\label{SFermionZ}
S_F^{(z)} & = & \int_0^\beta {\rm d} \tau \int {\rm d}^2 x \,
\Bigg\{
16 \left( t' {\tilde \rho}_1 + t'' {\tilde \rho}_2 \right)
\left(\partial_x \bar z \partial_x z 
+ \partial_y \bar z \partial_y z \right)
-\frac{2\tilde \rho \, \kappa}{a^2} \, 
\left(\bar z \, \partial_\tau z \right)^2
\nonumber \\ & & \qquad \qquad \qquad \quad
+ 16 \tilde \rho \left(J- t^2 \kappa_2 \right) 
\left( F^*_x F^{}_x + F^*_y F^{}_y \right)
- 8 \tilde \rho \left(J- 2 t^2 \kappa_1 \right)
\left( F^*_x F^{}_y + F^*_y F^{}_x \right)
\nonumber \\ & & \qquad \qquad \qquad \quad
+ 16 \tilde \rho \left( {t'}^2 \tilde \kappa_3 
+ 2 {t''}^2 \tilde \kappa_4 
+ 4 t' t'' \tilde \kappa_5 \right)
\left[ \left( \bar z \partial_x z \right)^2
+ \left( \bar z \partial_y z \right)^2 \right]
\Bigg\} \; , 
\eea
contains only contributions with the $z$-field, and
\bea
\label{SFermionZeta}
S_F^{(\zeta)} & = &  \int_0^\beta {\rm d} \tau \int {\rm d}^2 x \,
\Bigg\{
- \frac{4{\tilde \rho}}{a} 
\left( {\bar z} \partial_\tau \zeta 
+ {\bar \zeta} \partial_\tau z \right)
\nonumber \\ & & \qquad \qquad \qquad \quad
+ 8 \tilde \rho \left(J - 2 t^2 \kappa_1 \right)  
\left\{2 G^* G
+ \sqrt{2}
\left[
G^* \left( F_y -F_x \right) + \left( F^*_y -F^*_x \right) G
\right]\right\}
\Bigg\} \; ,
\eea
contains contributions with $\zeta$-fields.

\section{Integration of fast magnetic modes}
The effective action at this stage contains only the slow modes described by 
the $z$-fields and the fast modes corresponding to the $\zeta$-fields,
The next step is to integrate out the $\zeta$-fields. However, due to the
gauge freedom introduced by the CP$^1$ fields, a gauge fixing is necessary.

\subsection{Gauge fixing\label{GaugeFixing}}
In order to integrate out the $\zeta$-field we have to discuss 
the gauge-fixing for the $z$-fields and its consequences on the $\zeta$-fields.
As already mentioned in Sec.\ \ref{SlowCP1}, the condition (\ref{constzz})
can be used to fix a global phase of the $\zeta$-field with respect to the
one of the $z$-field. In order to see this, we can use the following 
parametrization of the $z$-field:
\bea
z & = &
\left[
\begin{array}{l}
\cos \left( \frac{\theta}{2} \right) \,
\exp \left\{ -i \left( \frac{\varphi}{2} - \Lambda \right) \right\} 
\\ \\
\sin \left(\frac{\theta}{2} \right) \,
\exp \left\{ +i \left( \frac{\varphi}{2} + \Lambda \right) \right\}
\end{array}
\right]
\; ,
\eea
where $0 \leq \theta \leq \pi$.
For $\zeta$ we use a similar parametrization
\bea
\zeta & = &
\left[
\begin{array}{l}
\rho_1 \,
\exp 
\left\{ -i \left( \frac{\chi}{2} - \Gamma \right) \right\} 
\\ \\
\rho_2 \,
\exp \left\{ +i \left( \frac{\chi}{2} + \Gamma \right) \right\}
\end{array}
\right]
\; .
\eea
Then, the condition (\ref{constzz}) can be fulfilled with 
$\Gamma = \Lambda + (2m+1) \pi/2$, with $m$ integer, that is equivalent to
\bea
\label{GaugeFixZeta}
\frac{\zeta_1}{\mid \zeta_1 \mid} \, \frac{\zeta_2}{\mid \zeta_2 \mid} & = &
{\rm e}^{2i \Gamma} = - {\rm e}^{2i \Lambda}
= -\frac{z_1}{\mid z_1 \mid} \, \frac{z_2}{\mid z_2 \mid} \; ,
\eea
that can be translated into
\bea
\label{zzAfterGauge}
1 - 
\frac{\mid z_1 \mid \mid \zeta_1 \mid}{\mid z_2 \mid \mid \zeta_2 \mid} = 0 
\; .
\eea
Then, the condition (\ref{constzz}) can be enforced as follows:
\bea
\label{constzzAfterChange}
\delta \left(\bar z \zeta + \bar \zeta z \right) & = & 
\frac{\mid z_2 \mid \mid \zeta_2 \mid}
{\mid z_1 \mid \left( \zeta^*_2 z^{}_2 + z^*_2 \zeta^{}_2 \right)} \;
\delta 
\left( \frac{\mid z_2 \mid \mid \zeta_2 \mid}{\mid z_1 \mid} - \mid \zeta_1 \mid
\right) \; ,
\eea
where we solved the constraint (\ref{zzAfterGauge}) in favor of 
$\mid \zeta_1 \mid$.

Once we enforced the constraint (\ref{constzz}), we impose the following 
gauge fixing for the $z$ field
\bea
\label{gaugeCP1}
z^{}_1 + z^*_1 = 0 \; ,
\eea
as normally done in a CP$^1$ theory. The implementation of the constraint
(\ref{constzzAfterChange}) in the part of the action containing $\zeta$-fields
(\ref{SFermionZeta}) is given in Appendix \ref{AfterConstraintZZeta}.
\subsection{Change of measure}
We discuss here the change of variables from the real vectors 
${\vec \Omega}^{A,B}$ to the complex fields $z$ and $\zeta$, taking into 
account the Jacobian of the transformation, together with the constraints 
for the $z$- and $\zeta$-fields already discussed in Sec.\ \ref{GaugeFixing}.
For this purpose we first consider (\ref{Omega-cp1}) and (\ref{cp1stag}).
\bea
\label{OmegaA}
{\vec \Omega}^A_j & = & \left( {\bar z}_j 
\sqrt{1 - a^2 {\bar \zeta}_j \zeta_j} - a {\bar \zeta}_j \right) \vec \sigma 
\left( z_j \sqrt{1 - a^2 {\bar \zeta}_j \zeta_j} - a \zeta_j \right)
\; ,
\\
\label{OmegaB}
{\vec \Omega}^B_j & = & 
\left( z_j \sqrt{1 - a^2 {\bar \zeta}_j \zeta_j} + a \zeta_j \right)
\, \sigma^y \, \vec \sigma \, \sigma^y \,
\left( z^*_j 
\sqrt{1 - a^2 {\bar \zeta}_j \zeta_j} + a \zeta^*_j \right) 
\; .
\eea
Then, we have
\bea
\mid {\vec \Omega}^A_j \mid^2 & = & 
\bigg[
{\bar z}_j  z_j \left(1 - a^2 {\bar \zeta}_j \zeta_j\right) 
+ a^2 {\bar \zeta}_j \zeta_j
-a \left( {\bar z}_j \zeta_j + {\bar \zeta}_j z_j \right)
\sqrt{1 - a^2 {\bar \zeta}_j \zeta_j}
\bigg]^2 \; ,
\nonumber \\
\mid {\vec \Omega}^B_j \mid^2 & = & 
\bigg[
{\bar z}_j  z_j \left(1 - a^2 {\bar \zeta}_j \zeta_j\right) 
+ a^2 {\bar \zeta}_j \zeta_j
+a \left( {\bar z}_j \zeta_j + {\bar \zeta}_j z_j \right)
\sqrt{1 - a^2 {\bar \zeta}_j \zeta_j}
\bigg]^2 \; .
\eea
Therefore,
\bea
\label{deltas}
\delta \left( \mid {\vec \Omega}^A \mid -1 \right) \,
\delta \left( \mid {\vec \Omega}^B \mid -1 \right) 
\simeq
\left( 1 + \frac{3}{2} a^2 \bar \zeta \zeta \right) \,
\delta \left( \bar z z -1 \right) \, 
\delta \left( \bar \zeta z + \bar z \zeta \right) \; .
\eea
The transformation of the last constraint due to gauge fixing was 
already discussed in (\ref{constzzAfterChange}). On the other hand, 
we have due to (\ref{zzAfterGauge})
\bea
\label{barzetazeta}
\bar \zeta \zeta & = & 
\frac{\mid \zeta_2 \mid^2}{\mid z_1 \mid^2} \; ,
\eea
where we used $\bar z z = 1$.

Due to gauge fixing, and without imposing the constraints on the moduli
of the vector fields ${\vec \Omega}^A$ and ${\vec \Omega}^B$, since 
we take them into account with (\ref{deltas}), we have
independent variables 
${\rm Im} z_1$, ${\rm Re} z_2$, ${\rm Im} z_2$, $\mid \zeta_1 \mid$,
${\rm Re} \zeta_2$, and ${\rm Im} \zeta_2$. 
The Jacobian going from vector fields to CP$^1$ variables is in an expansion
in powers of $a$,  
\bea
{\cal J} = a^3 \left( {\cal J}_0 + a^2 {\cal J}_2 \right) + 
{\cal O} \left(a^6\right) \; ,
\eea
where 
\bea
{\cal J}_0 = 32 \,
\frac{i \mbox{Im} z_1 \left( 2 \mid z_1 \mid^2 -1 \right)}
{\mid z_1 \mid} \, 
\frac{\left( \zeta^*_2 z^{}_2 + z^*_2 \zeta^{}_2 \right)}
{\mid z_2 \mid \mid \zeta_2 \mid} \; ,
\eea
and
\bea
{\cal J}_2 = -48  \,
\frac{i \mbox{Im} z_1 \left( 2 \mid z_1 \mid^2 -1 \right)}{\mid z_1 \mid} \, 
\frac{\left( \zeta^*_2 z^{}_2 + z^*_2 \zeta^{}_2 \right) \mid \zeta_2 \mid}
{\mid z_2 \mid} \; ,
\eea
such that we finally have
\bea
{\cal J} = a^3 32 \,
\frac{i \mbox{Im} z_1 \left( 2 \mid z_1 \mid^2 -1 \right)}
{\mid z_1 \mid} \, 
\frac{\left( \zeta^*_2 z^{}_2 + z^*_2 \zeta^{}_2 \right)}
{\mid z_2 \mid \mid \zeta_2 \mid} 
\left( 1 - \frac{3}{2} a^2 \frac{\mid \zeta_2 \mid^2 }{\mid z_1 \mid^2} \right)
\; .
\eea

Taking into account the Jacobian above, the constraints in Sec.\ 
\ref{GaugeFixing}, the transformation of the constraints (\ref{deltas})
together with (\ref{barzetazeta}), the measure changes as follows
\bea
\label{measure}
\int {\rm d} {\vec \Omega}^A \, {\rm d} {\vec \Omega}^B 
\; \delta \left( \mid {\vec \Omega}^A \mid -1 \right)
\; \delta \left( \mid {\vec \Omega}^B \mid -1 \right)
& \rightarrow & 
\int {\rm d} {\rm Im} z_1 \, {\rm d} {\rm Re} z_2 \, {\rm d} {\rm Im} z_2 \, 
\delta \left( \bar z z -1 \right) \,
\frac{i \mbox{Im} z_1 \left( 2 \mid z_1 \mid^2 -1 \right)}
{\mid z_1 \mid^2}
\nonumber \\ & & \qquad \times
\int {\rm d} {\rm Re} \zeta_2 \, {\rm d} {\rm Im} \zeta_2
\; .
\eea

\subsection{Effective field theory for the magnetic properties of the t-J model}
The final step is to integrate out the $\zeta$-field. Therefore, we 
concentrate on the corresponding integrals.
\bea
\label{IntegralOverZeta}
\lefteqn{
\int {\cal D} {\rm Re} \zeta_2 \, {\cal D} {\rm Im} \zeta_2 \,
{\rm e}^{-S_{\rm eff}^{(\zeta)}} 
}
\nonumber \\ & & =
\int {\cal D} {\rm Re} \zeta_2 \, {\cal D} {\rm Im} \zeta_2 \,
\exp \left\{
-\int_0^\beta {\rm d} \tau \int {\rm d}^2 x \, \left[
\zeta_2^* \Delta \zeta_2^{} + \Lambda^* \zeta^2_2 + \Lambda \zeta^{*2}_2
+ \left(\Xi -\Gamma \right) \zeta^*_2 
- \left(\Xi^* + \Gamma^* \right) \zeta^{}_2 
\right]
\right\}
\; ,
\eea
where we introduced the following notation:
\bea
\label{DefDeltaLambdaXiGamma}
\tilde J & = & \left[J+8 \tilde \rho \left(J - 2 t^2 \kappa_1 \right) \right]
\; ,
\nonumber \\
\Delta & = & 8 \tilde J
\frac{\mid z_2 \mid^4+\mid z_1 \mid^4}{\mid z_1 \mid^2} \; ,
\nonumber \\
\Lambda & = & 8 \tilde J
z_2^2 \; ,
\nonumber \\ 
\Xi & = & 
\left[\frac{2\left(1-2 \tilde \rho\right)}{a} \right] z_2
\left(\frac{1}{\mid z_1 \mid^2} \, z^{}_1 \, \partial_\tau z_1^*
+ \frac{1}{\mid z_2 \mid^2} \, z^*_2 \, \partial_\tau z^{}_2 \right) \; ,
\nonumber \\
\Gamma & = &  
2 \sqrt{2} \tilde J
z_2 
\left[\frac{z^*_1 z^*_2 }{\mid z_2 \mid^2} \left( F_y -F_x \right)
+ \left( F^*_y -F^*_x \right)
\frac{z^{}_1z^{}_2}{\mid z_1 \mid^2}
\right] 
.
\eea

We go now over to real and imaginary parts $\zeta_2 \equiv \eta + i \xi$, 
such that for eq.\ (\ref{IntegralOverZeta}) we have
\bea
\label{IntegralOverZeta''}
\int {\cal D} {\rm Re} \zeta_2 \, {\cal D} {\rm Im} \zeta_2 \,
{\rm e}^{-S_{\rm eff}^{(\zeta)}} 
\propto 
\int {\cal D} \eta \, {\cal D} \xi 
\exp \left[
-\int_0^\beta {\rm d} \tau \int {\rm d}^2 x \, \left(
{\cal A} \eta^2 + {\cal B} \xi^2 + 2i {\cal C} \eta \xi + {\cal D} \eta 
+ i {\cal E} \xi
\right)
\right] \; ,
\eea
with ${\cal A} \equiv \Delta + \Lambda^* + \Lambda$, 
${\cal B} \equiv \Delta - \left(\Lambda^* + \Lambda \right)$,
${\cal C} \equiv \Lambda^* - \Lambda$,
${\cal D} \equiv \Xi - \Xi^* - \left(\Gamma^* + \Gamma \right)$,
and 
${\cal E} \equiv -\left(\Xi + \Xi^*\right) + \left(\Gamma - \Gamma^* \right)$.
Performing the integrals over the $\zeta$-fields leads to the effective action
\bea
\label{EffectiveActionAfterIntegrationOfZeta}
{\tilde S}_{eff} & = & \int {\rm d} \tau \sum_j \Bigg\{ 
\frac{1}{2} \ln \left({\cal A}{\cal B} + {\cal C}^2 \right)
+ \frac{\left({\cal A}{\cal E}^2 - 2 {\cal C}{\cal D}{\cal E} 
- {\cal B} {\cal D}^2 \right)}
{4 \left({\cal A}{\cal B} + {\cal C}^2\right)}
\Bigg\} 
\; .
\eea

Next we proceed to evaluate the different contributions to the action.
We consider first the term that affects the measure
\bea
\label{AB+C}
{\cal A}{\cal B} + {\cal C}^2 & = & 
\left\{
\frac{8}{\mid z_1 \mid^2} 
\left[J+8 \tilde \rho \left(J - 2 t^2 \kappa_1 \right) \right]
\left(1-2 \mid z_1\mid^2\right) 
\right\}^2 \; .
\eea
Then,
\bea
\exp \left[ - \frac{1}{2} \ln \left({\cal A}{\cal B} 
+ {\cal C}^2 \right) \right]
& = & \left\{
\frac{8}{\mid z_1 \mid^2} 
\left[J+8 \tilde \rho \left(J - 2 t^2 \kappa_1 \right) \right]
\left(1-2 \mid z_1\mid^2\right) 
\right\}^{-1} \; ,
\eea
such that the measure and constraints of the path-integral are now
\bea
Z & \rightarrow & \int {\cal D} \bar z \, {\cal D} z \,
\delta \left( \bar z z -1 \right) \,
\delta \left( z_1 + z_1^* \right) \, \left( z^{}_1 - z^*_1 \right) \cdots \; ,
\eea
that are the ones usually appearing in the CP$^1$ model.

After a lengthy but straightforward calculation we have
\bea
\frac{\left({\cal A}{\cal E}^2 - 2 {\cal C}{\cal D}{\cal E} 
- {\cal B} {\cal D}^2 \right)}
{4 \left({\cal A}{\cal B} + {\cal C}^2\right)} & = &
\Bigg\{ \frac{1}{2 \tilde J} 
\left[\frac{\left(1-2 \tilde \rho\right)}{a} \right]^2
\left[\partial_\tau \bar z \, \partial_\tau z 
+ \left( \bar z \partial_\tau z \right)^2 \right] 
\nonumber \\ & & \quad
+ \frac{1}{\sqrt{2}}\left[\frac{\left(1-2 \tilde \rho\right)}{a} \right]
\left[ \partial_\tau \bar z \left(\partial_y z - \partial_x z \right) 
- \partial_\tau z \left(\partial_y \bar z - \partial_x \bar z\right)\right]
\nonumber \\ & & \quad
- \tilde J \bigg[
\partial_x \bar z \partial_x z 
+ \partial_y \bar z \partial_y z 
+ \left( \bar z \partial_x z \right)^2 
+ \left( \bar z \partial_y z \right)^2  
\nonumber \\ & & \qquad \quad
- \partial_x \bar z \partial_y z - \partial_y \bar z \partial_x z  
- 2 \left( \bar z \partial_x z \right) 
\left( \bar z \partial_y z \right)
\bigg]
\Bigg\} \; .
\eea

Taking into account the contribution to the action in 
(\ref{SFermionZ}), that 
remains unaffected by the integration over the $\zeta$-fields, 
we arrive at the effective 
action for the $z$-fields
\bea
\label{FinalAction}
S & = & \int {\rm d} \tau \, {\rm d}^2 x \sum_\mu \frac{1}{g_\mu}
\left[ \partial_\mu \bar z \, \partial_\mu z 
+ \gamma_\mu \left( \bar z \partial_\mu z \right)^2 \right]
\; ,
\eea
with $\mu = \tau$, $x$, $y$, and
\bea
\label{FinalCouplings}
g_\tau & = & \frac{2 \tilde Ja^2}{\left(1-2 \tilde \rho\right)^2} \; ,
\nonumber \\
g_x = g_y & = & \left[ J \left( 1 + 8 \tilde \rho \right) 
+ 16 \tilde \rho t^2 \left(\kappa_1 - \kappa_2 \right) 
+ 16 \left( t' {\tilde \rho}_1 + t'' {\tilde \rho}_2 \right)\right]^{-1} \; ,
\nonumber \\
\gamma_\tau & = & 1 - 
\frac{4 \tilde \rho \kappa \tilde J}{\left(1-2 \tilde \rho\right)^2} \; ,
\nonumber \\
\gamma_x = \gamma_y & = & 
\frac{
J \left( 1 + 8 \tilde \rho \right) 
+ 16 \tilde \rho \left[t^2 \left(\kappa_1 - \kappa_2 \right) 
+ {t'}^2 \kappa_3 + 2 {t''}^2 \kappa_4 + 4 t' t'' \kappa_5 \right]
}{\left[ J \left( 1 + 8 \tilde \rho \right) 
+ 16 \tilde \rho t^2 \left(\kappa_1 - \kappa_2 \right) 
+ 16 \left( t' {\tilde \rho}_1 + t'' {\tilde \rho}_2 \right)\right]},
\eea
where $\tilde J$ was defined in (\ref{DefDeltaLambdaXiGamma}). 
Equation (\ref{FinalAction}) together with (\ref{FinalCouplings})
are the main result of this work.

\section{Discussion of the results and
conclusions \label{conclusion}}
The action (\ref{FinalAction}) at which we arrived, is of the form previously 
analyzed in the context of frustrated quantum antiferromagnets 
\cite{chubukov94,azaria95,chubukov96}. In the absence of doping, i.e.\ setting 
$\tilde \rho = {\tilde \rho}_1 = {\tilde \rho}_2 = 0$, we have
$\gamma_\tau = \gamma_x = \gamma_y =1$, such that the model reduces to the
CP$^1$ model appropriate for a Heisenberg antiferromagnet, making thus 
explicitely the connection with the undoped case. As is well known, the
excitations in that case correspond to bosons describing 
the transverse fluctuations of a vector field pertaining to the 
O(3) non-linear 
$\sigma$-model. From the point of view of the CP$^1$ model this corresponds
to a phase where the $z$-fields are confined.   

However, in the presence of doping, when the couplings $\gamma_\mu \neq 1$,
the model does not correspond any more to a collinear antiferromagnet but
describes in general coplanar incommensurate quantum antiferromagnets
\cite{chubukov94,azaria95}. This can be seen by constructing $R \in$ SO(3)
out of matrices $g \in$ SU(2), as follows:
\bea
\label{FromSU2ToSO3}
R_{ab} (g) & = & \frac{1}{2} \mbox{Tr} \, \sigma^a g \sigma^b g^\dagger
\; ,
\eea 
where $g$ is given in terms of the $z$-fields like in (\ref{matrixU}). Then,
it can be readily shown that the action (\ref{FinalAction}) can be expressed
in terms of SO(3) fields:
\bea
\label{FinalActionInSO3}
S & = & \int {\rm d} \tau \, {\rm d} x^2 \sum_\mu \frac{1}{8g_\mu}
\mbox{Tr} \left( \partial_\mu R \partial_\mu R^{-1} - 
\gamma_\mu \partial_\mu R Q \partial_\mu R^{-1} \right) \; ,
\eea
where $Q = ¸\mbox{diag} (1,1,-1)$. Expressing the matrix $R$ as 
$R=({\vec n}_1,{\vec n}_2,{\vec n}_3)$, where the vectors ${\vec n}_i$
fulfill ${\vec n}_i^2 = 1$ for $i=1,2,3$ and 
${\vec n}_i \cdot {\vec n}_j = 0$ for $i \neq j$,
the action (\ref{FinalActionInSO3}) takes the form
\bea
S & = &
\int {\rm d} \tau \, {\rm d} x^2 \sum_\mu \frac{1}{4g_\mu}
\left[ \sum_{i=1}^2
\partial_\mu {\vec n}_i \cdot \partial_\mu {\vec n}_i 
- \left(1+\gamma_\mu \right) 
\left( {\vec n}_1 \cdot \partial_\mu {\vec n}_2 \right)^2 
\right] \; ,
\eea
showing that in general a coplanar configuration is favored. In the case 
$\gamma_\mu = 1$,
the action above reduces as expected to the O(3) non-linear $\sigma$-model
by virtue of the relation
\bea
\partial_\mu {\vec n}_3 \cdot \partial_\mu {\vec n}_3 & = &
\partial_\mu {\vec n}_1 \cdot \partial_\mu {\vec n}_1 +
\partial_\mu {\vec n}_2 \cdot \partial_\mu {\vec n}_2 
- 2 \left( {\vec n}_1 \cdot \partial_\mu {\vec n}_2 \right)^2 \; .
\eea

On the other hand, for the general case $\gamma_\mu \neq 1$, assuming for
simplicity that $\gamma_\mu = \gamma$ and $g_\mu = g$ for 
$\mu = \tau, x, y$, it can be seen that, following Ref.\ \cite{azaria95}, the 
gauge fields responsible for confinement in the case 
$\gamma = 1$, acquire a mass 
\bea
M^2 = -\frac{2}{g} \left( 1 - \frac{1}{\gamma} \right) \; .
\eea 
Moreover, it is expected that in the
infrared limit, the behavior is dominated by an O(4) fixed point
\cite{chubukov94,azaria95}, i.e.\ 
it is expected that the coupling $\gamma$ scales to zero, raising the 
mass of the gauge fields. Further arguments in Cavour of deconfinement were 
advanced by showing that for $\gamma \neq 1$, the mass
of the gauge field is inversely proportional to the square root of the 
temperature \cite{chubukov96}, 
and therefore, a confinement-deconfinement transition as a function of 
temperature can be expected when $\gamma$ departs from 1. 
Hence, we see that the possibility of deconfined
(bosonic) spinons is opened due to doping of a collinear antiferromagnet. 

Finally, we would like to remark, that the present results show some  
differences from those obtained previously by one of the authors 
\cite{muramatsu90} and collaborators, dealing with a doped antiferromagnet 
modeled by the
so-called spin-fermion model. This model is the starting point that leads
to the $t$-$J$ model in the limit where the exchange coupling between the 
dopant hole, that mainly resides on the oxygen orbitals, and the copper hole
form the Zhang-Rice singlet \cite{zhang88}. The gradient expansion for
the spin-fermion model based on the assumption of a short-range 
antiferromagnetic order led to an O(3) non-linear $\sigma$-model,
that as a function of doping had a transition to the corresponding
quantum disordered phase \cite{kuebert95}. Hence, under the same assumption
as in the present work, no hint to deconfinement of spinons was obtained.  
An SO(3) non-linear $\sigma$-model can be obtained from the spin-fermion model
only under the assumption of an incommensurate coplanar short-range order
in the microscopic model \cite{klee96}. Therefore, although considering
the symmetries present in a doped antiferromagnet, both models appear 
equivalent, we conclude on the basis of
the present results that imposing the constraint against
double occupancy leads to a richer picture of the possible phases of a
doped antiferromagnet, with a minimal number of assumptions.   

\vspace*{0.5cm}
\par
\noindent
{\bf Acknowledgments}
We are grateful to A.\ Greco, A.\ Dobry and A.\ Foussats for interesting
discussions in the early stages of this work. We thank also M.\ Barbosa
da Silva Neto for illuminating discussions on incommensurate coplanar phases 
and a critical reading of the manuscript.

\appendix

\section{Set of constraints for the $t$-$J$ model \label{ListOfConstraints}}
We give here the explicit expressions of the constraints arising from 
considering the canonical momenta of the system, and the matrix of constraints.
First we recall the definitions of right and left derivatives in order to
reach a self-contained presentation. 

Given generators $z^a$ of a Berezin algebra
with $a=1,\dots,k$, the right derivative is defined as
\bea
\frac{\partial_r}{\partial z^a} 
z^{a_1} \cdots z^{a_k} & = &
\sum_{i=1}^k \left(-1\right)^{\sum_{j=k}^{i+1} P_{(a)} P_{(a_j)}}
\delta_{a,a_1} \, z^{a_1} \cdots z^{a_{i-1}} z^{a_{i+1}} \cdots  z^{a_k}
\; ,
\eea
where $P_{(a)} = 0,1$ depending on the parity of the generator (even or odd,
respectively). For later use, we introduce also left derivatives defined as
\bea
\frac{\partial_\ell}{\partial z^a} 
z^{a_1} \cdots z^{a_k} & = &
\sum_{i=1}^k \left(-1\right)^{\sum_{j=1}^{i-1} P_{(a)} P_{(a_j)}}
\delta_{a,a_1} \, z^{a_1} \cdots z^{a_{i-1}} z^{a_{i+1}} \cdots  z^{a_k}
\; .
\eea
In both cases it is understood that when the index $j$ runs beyond the
interval $[1,k]$, then $P_{(a_j)} = 0$. Both derivatives are equivalent to 
ordinary derivatives when dealing with even generators $z^a$.

With the rules above we obtain the following constraints from 
(\ref{CanonicalMomenta}).
\bea
\begin{array}{lcllcllcl}
\phi^{(3)} & = & \Pi^{++} \; , &
\phi^{(4)} & = & \Pi^{+-} +
i \frac{\left(1+\rho \right) u -1}
{\left(2-v\right)^2 - 4 \rho -u^2} \, 
X^{-+} \; , &
\phi^{(5)} & = & \Pi^{-+} -
i \frac{\left(1+\rho \right) u -1}
{\left(2-v\right)^2 - 4 \rho -u^2} \, 
X^{+-} \; , \\
\phi^{(6)} & = & \Pi^{--} \; , &
\phi^{(8)} & = & \Pi^{00} \; , &
\phi^{(11)} & = & \Pi^{+0} - \frac{i}{2}  X^{0 +} \; ,\\
\phi^{(12)} & = & \Pi^{-0} - \frac{i}{2}  X^{0 -} \; , &
\phi^{(13)}& = & \Pi^{0+} - \frac{i}{2}  X^{+0} \; , &
\phi^{(14)} & = & \Pi^{0-} - \frac{i}{2}  X^{-0} \; .
\end{array}
\eea

The matrix of constraints (\ref{MatrixOfConstraints}) has the following 
components
\bea
{\vec A}^{(0)} 
& = & 
\left(
\begin{array}{cccccccc}
0 & 0 & 1 & 0 & 0 & 1 & 0 & 0 \\
0 & 0 & 1 - X^{--} &  X^{-+} & X^{+-} & 1 - X^{++} & 0 & 0\\
-1 & X^{--} - 1 & 0 & \frac{i}{4 X^{+-}} & -\frac{i}{4 X^{-+}} & 0 & 0 & 0\\
0 & -X^{-+} & -\frac{i}{4 X^{+-}} & 0 & 
-\frac{i}{X^{++}} &
-\frac{i\left(1+X^{--}\right)}{4 X^{++} X^{+-}} & 0 & 0\\
0 & -X^{+-} &  -\frac{i}{X^{+-}} &  
\frac{i}{X^{++}} & 0 &
\frac{i\left(1+X^{--}\right)}{4 X^{++} X^{-+}} & 0 & 0\\
-1 & X^{++}-1 & 0 &
\frac{i\left(1+X^{--}\right)}{4 X^{++} X^{+-}}
& -\frac{i\left(1+X^{--}\right)}{4 X^{++} X^{-+}} & 0 & 0 & 0 \\
0 & 0 & 0 & 0 & 0 & 0 & 0 & 1 \\
0 & 0 & 0 & 0 & 0 & 0 & -1 & 0 
\end{array}
\right) \; ,
\\
{\vec A}^{(1)} & = &
\left(
\begin{array}{cccccccc}
0 & 0 & 0 & 0 & 0 & 0 & 0 & 0\\
0 & 0 & 0 & 0 & 0 & 0 & 0 & 0\\
0 & 0 & 0 & 
\frac{i\left(2+X^{++} \right) \rho}{X^{++} X^{+-}} &
-\frac{i\left(2+X^{++} \right) \rho}{X^{++} X^{-+}} & 0 & 0 & 0\\
0 & 0 & 
-\frac{i\left(2+X^{++} \right) \rho}{X^{++} X^{+-}} & 0 &
-\frac{i\rho}{X^{++}} & 
-\frac{i\left(2+ X^{--}\right)\rho}{4X^{++} X^{+-}} & 0 & 0\\
0 & 0 & \frac{i\left(2+X^{++} \right) \rho}{X^{++} X^{-+}}  &
\frac{i\rho}{X^{++}} & 0 &
\frac{i\left(2+ X^{--} \right)\rho}{4X^{++} X^{-+}} & 0 & 0\\
0 & 0 & 0 & \frac{i\left(2+ X^{--} \right)\rho}{4X^{++} X^{+-}} &
-\frac{i\left(2+ X^{--}\right)\rho}{4X^{++} X^{-+}} & 0 & 0  & 0 \\
0 & 0 & 0 & 0 & 0 & 0 & 0 & 0 \\
0 & 0 & 0 & 0 & 0 & 0 & 0 & 0 
\end{array}
\right) \; .
\eea
The matrix $\vec B$ contains only single Grassmann generators. 
\bea
\vec B & = & \left(
\begin{array}{cccccc}
0 & 0 & X^{0+} & X^{0-} & -X^{+0} & -X^{-0} \\
0 & 0 & X^{0+} & X^{0-} & -X^{+0} & -X^{-0} \\
B_{31} & X^{-0} & 0 & 0 & 0 & 0 \\
B_{41} & 0 & B_{43} & B_{44} & B_{45} & B_{46} \\
0 & -X^{+0} & B_{53} & B_{54} & B_{55} & B_{56} \\
0 & 0 & 0 & 0 & 0 & 0 \\
0 & 0 & -X^{0+} & -X^{0-} & X^{+0} &  X^{-0} \\
0 & 0 & 0 & 0 & 0 & 0
\end{array}
\right) \; ,
\eea
where
\bea
\begin{array}{rclrcl}
{\vec B}_{31} & = & \frac{X^{0+} X^{+-}}{\left(X^{++}\right)^2} \; , &
{\vec B}_{41} & = & - \frac{X^{0+}}{X^{++}} \; , \\
{\vec B}_{43} & = & 
-\frac{i\left(1+X^{--} + 2 X^{++} X^{--} \right)X^{0+}}{4X^{++} X^{+-}} \; ,
&
{\vec B}_{44} & = & 
-\frac{i\left(1+X^{--} + 2 X^{++} X^{--} \right)X^{0-}}{4X^{++} X^{+-}} \; ,
\\
{\vec B}_{45} & = &  
\frac{i\left(1+X^{--} + 2 X^{++} X^{--} \right)X^{+0}}{4X^{++} X^{+-}} \; ,
&
{\vec B}_{46} & = & 
\frac{i\left(1+X^{--}+2 X^{++} X^{--}\right)X^{-0}}{4X^{++} X^{--} X^{+-}} \; ,
\\ 
{\vec B}_{53} & = &  
\frac{i\left(1+X^{--}+2 X^{++} X^{--} \right)X^{0+}}{4X^{++} X^{-+}} \; ,
& 
{\vec B}_{54} & = & 
\frac{i\left(1+X^{--}+2 X^{++} X^{--} \right)X^{0-}}{4X^{++} X^{-+}} \; ,
\\
{\vec B}_{55} & = &  
-\frac{i\left(1+X^{--}+2 X^{++} X^{--} \right)X^{+0}}{4X^{++} X^{-+}} \; ,
&
{\vec B}_{56} & = & 
-\frac{i\left(1+X^{--}+2 X^{++} X^{--} \right)X^{-0}}{4X^{++} X^{-+}} \; .
\end{array}
\eea
Furthermore, $\vec C = -{\vec B}^T$, and
\bea
{\vec D} & = &
\left(
\begin{array}{cccccc}
0 & 0 & 0 & 0 & \frac{X^{+-}}{X^{++}} & -1 \\
0 & 0 & X^{-+} & -X^{++} & 0 & 0 \\
0 & X^{-+} & 0 & 0 & -i & 0 \\
0 & -X^{++} & 0 & 0 & 0 & -i \\
\frac{X^{+-}}{X^{++}} & 0 & -i & 0 & 0 & 0 \\
-1 & 0 & 0 & -i & 0 & 0 
\end{array}
\right) \; .
\eea

\section{Fermionic terms in the staggered CP$^1$ representation\label{HoppingCP1}}
We give here the explicit expressions for the fermionic contributions up
to ${\cal O} \left(a^2\right)$. Defining vectors 
\bea
\begin{array}{rclrclrclrclrcl}
{\vec x}^{(1)} & = & \left(0,0\right) , &
{\vec x}^{(2)} & = & \left( 0, \sqrt{2} a \right) , &
{\vec x}^{(3)} & = & \sqrt{2} a \left( -1, 1 \right) , &
{\vec x}^{(4)} & = & \sqrt{2} a \left( -1, 0 \right) , &
\end{array}
\eea
we have for the fermionic terms $\Xi^{(i)}_e ({\vec k}_1, {\vec k}_2 )$ 
with $i= 1,\dots,4$
\bea
\Xi^{(i)}_{AB} ({\vec k}_1,{\vec k}_2) & = & -\frac{2a}{N} 
\exp \left\{ i \left[ {\vec k}_1 \cdot {\vec x}_A -
{\vec k}_2 \cdot \left({\vec x}_B + {\vec x}^{(i)} \right)\right] \right\} 
\sum_j 
\exp \left[ i \left({\vec k}_1 - {\vec k}_2 \right) \cdot {\vec x}_j \right] 
{\tilde U}_{j,AB}^{(i)} \; ,
\eea
where ${\tilde U}_{j,AB}^{(i)}$ denotes the matrix element of the product of 
the SU(2) matrices in nearest neighbor sites, connecting sublattices $A$ and 
$B$ within the unit cell $j$. The index $i$ denotes the four nearest neighbors 
starting from sublattice $A$.
Using the notation introduced in (\ref{DefGAndF}), the gradient 
expansion of the products of the SU(2) matrices up to ${\cal O}(a)$
have the following form 
\bea
\label{xi0}
\begin{array}{rclrcl}
{\tilde U}_{j,AB}^{(1)} & = &
G_j^* \; ,
&
{\tilde U}_{j,AB}^{(2)} & = &
G_j^* + \sqrt{2}  F_{jy}^*
\; ,
\\
{\tilde U}_{j,AB}^{(3)} & = &
G_j^* - \sqrt{2}  \left(F_{jx}^* - F_{jy}^* \right)
\; ,
&
{\tilde U}_{j,AB}^{(4)} & = &
G_j^* - \sqrt{2} F_{jx}^*
\; .
\end{array}
\eea
For the fermionic terms $\Xi^{(i)}_{BA} ({\vec k}_1, {\vec k}_2 )$ we have
\bea
\Xi^{(i)}_{BA} ({\vec k}_1,{\vec k}_2) & = & -\frac{2a}{N} 
\exp \left\{ i \left[ {\vec k}_1 \cdot {\vec x}_B -
{\vec k}_2 \cdot \left({\vec x}_A - {\vec x}_i \right)\right] \right\} 
\sum_j 
\exp \left[ i \left({\vec k}_1 - {\vec k}_2 \right) \cdot {\vec x}_j \right] 
{\tilde U}_{j,BA}^{(i)} \; ,
\eea
with ${\tilde U}_{j,BA}^{(i)} = {\tilde U}_{j,AB}^{(i)*}$. 

Next we list the terms originating from contributions proportional to $t'$. 
For $\Psi^{(i)}_{AA} ({\vec k}_1, {\vec k}_2 )$, $i=1,\dots,4$, we have
\bea
\label{psie}
\Psi^{(1,3)}_{AA} ({\vec k}_1, {\vec k}_2 ) & = & \frac{2}{N} 
\exp \left[ \pm i {\vec k}_2 \cdot {\vec x}_4 \right]
\sum_j 
\exp \left[ i \left({\vec k}_1 - {\vec k}_2 \right) \cdot {\vec x}_j \right] 
{\tilde U}_{j,AA}^{(1,3)} 
\; ,
\\
\Psi^{(2,4)}_{AA} ({\vec k}_1, {\vec k}_2 ) & = & \frac{2}{N} 
\exp \left[ \mp i {\vec k}_2 \cdot {\vec x}_2 \right]
\sum_j 
\exp \left[ i \left({\vec k}_1 - {\vec k}_2 \right) \cdot {\vec x}_j \right] 
{\tilde U}_{j,AA}^{(2,4)} 
\; ,
\eea 
where ${\tilde U}^{(i)}_{j,AA}$ denotes the product of SU(2) matrices on 
second nearest neighbor sites on sublattice $A$ along the diagonals, 
starting from the unit cell $j$, and $i$ numbers the four possibilities.
They have the following expansion in powers of $a$:
\bea
{\tilde U}_{j,AA}^{(1)} & = &
\sqrt{2} a {\bar z}_j \partial_x z_j 
+ a^2 \left[{\bar z}_j \partial^2_x z_j - \sqrt{2} 
\left( {\bar z}_j \partial_x \zeta_j + {\bar \zeta}_j \partial_x z_j
\right)\right] \; ,
\nonumber \\
{\tilde U}_{j,AA}^{(2)} & = &
\sqrt{2} a {\bar z}_j \partial_y z_j 
+ a^2 \left[{\bar z}_j \partial^2_y z_j - \sqrt{2} 
\left( {\bar z}_j \partial_y \zeta_j + {\bar \zeta}_j \partial_y z_j
\right)\right] \; ,
\nonumber \\
{\tilde U}_{j,AA}^{(3)} & = &
- \sqrt{2} a {\bar z}_j \partial_x z_j 
+ a^2 \left[{\bar z}_j \partial^2_x z_j + \sqrt{2} 
\left( {\bar z}_j \partial_x \zeta_j + {\bar \zeta}_j \partial_x z_j
\right)\right] \; ,
\nonumber \\
{\tilde U}_{j,AA}^{(4)} & = &
- \sqrt{2} a {\bar z}_j \partial_y z_j 
+ a^2 \left[{\bar z}_j \partial^2_y z_j + \sqrt{2} 
\left( {\bar z}_j \partial_y \zeta_j + {\bar \zeta}_j \partial_y z_j
\right)\right] \; .
\eea
For $\Psi^{(i)}_{BB}$, we have the same phase factors as for 
$\Psi^{(i)}_{AA}$ but we have to replace 
${\tilde U}_{j,AA}^{(i)}$ by ${\tilde U}_{j,BB}^{(i)}$ that are as follows:
\bea
{\tilde U}_{j,BB}^{(1)} & = &
\sqrt{2} a z_j \partial_x {\bar z}_j 
+ a^2 \left[z_j \partial^2_x {\bar z}_j + \sqrt{2} 
\left( z_j \partial_x {\bar \zeta}_j + \zeta_j \partial_x {\bar z}_j
\right)\right] \; ,
\nonumber \\
{\tilde U}_{j,BB}^{(2)} & = &
\sqrt{2} a z_j \partial_y {\bar z}_j 
+ a^2 \left[z_j \partial^2_y {\bar z}_j + \sqrt{2} 
\left( z_j \partial_y {\bar \zeta}_j + \zeta_j \partial_y {\bar z}_j
\right)\right] \; ,
\nonumber \\
{\tilde U}_{j,BB}^{(3)} & = &
- \sqrt{2} a z_j \partial_x {\bar z}_j 
+ a^2 \left[z_j \partial^2_x {\bar z}_j - \sqrt{2} 
\left( z_j \partial_x {\bar \zeta}_j + \zeta_j \partial_x {\bar z}_j
\right)\right] \; ,
\nonumber \\
{\tilde U}_{j,BB}^{(4)} & = &
- \sqrt{2} a z_j \partial_y {\bar z}_j 
+ a^2 \left[z_j \partial^2_y {\bar z}_j - \sqrt{2} 
\left( z_j \partial_y {\bar \zeta}_j + \zeta_j \partial_y {\bar z}_j
\right)\right] \; .
\eea

For the contributions proportional to $t''$ we have for processes involving 
the sublattice $A$,
\bea
\label{phi1e}
\Phi^{(1,3)}_{AA} ({\vec k}_1, {\vec k}_2 ) & = & \frac{2}{N} 
\exp \left[ \pm i {\vec k}_2 \cdot {\vec x}_3 \right]
\sum_j 
\exp \left[ i \left({\vec k}_1 - {\vec k}_2 \right) \cdot {\vec x}_j \right] 
{\bar U}_{j,AA}^{(1,3)} 
\; ,
\\
\Phi^{(2,4)}_{AA} ({\vec k}_1, {\vec k}_2 ) & = & \frac{2}{N} 
\exp \left[ \mp i {\vec k}_2 \cdot \left({\vec x}_2 - {\vec x}_4 \right)\right]
\sum_j 
\exp \left[ i \left({\vec k}_1 - {\vec k}_2 \right) \cdot {\vec x}_j \right]
{\bar U}_{j,AA}^{(2,4)} 
\; ,
\eea
where ${\bar U}^{(i)}_{j,AA}$ denotes the product of SU(2) matrices on 
second nearest neighbor sites on sublattice $A$ along the principal axes, 
starting from the unit cell $j$, and $i$ numbers the four possibilities.
They have the following expansion in powers of $a$:
\bea
{\bar U}_{j,AA}^{(1)} & = & \sqrt{2} a \left({\bar z}_j \partial_x z_j 
- {\bar z}_j \partial_y z_j\right)
\nonumber \\ & &
+ a^2 \left\{
{\bar z}_j \partial^2_x z_j - 2 {\bar z}_j \partial_{xy} z_j 
+ {\bar z}_j \partial^2_y z_j
-  \sqrt{2} \left[ {\bar z}_j \left(\partial_x \zeta_j 
- \partial_y \zeta_j \right) 
+ {\bar \zeta}_j \left(\partial_x z_j - \partial_y z_j \right)
\right]\right\} \; ,
\nonumber \\
{\bar U}_{j,AA}^{(2)} & = & \sqrt{2} a \left({\bar z}_j \partial_x z_j 
+ {\bar z}_j \partial_y z_j\right)
\nonumber \\ & &
+ a^2 \left\{
{\bar z}_j \partial^2_x z_j + 2 {\bar z}_j \partial_{xy} z_j 
+ {\bar z}_j \partial^2_y z_j
-  \sqrt{2} 
\left[ {\bar z}_j \left(\partial_x \zeta_j 
+ \partial_y \zeta_j \right) 
+ {\bar \zeta}_j \left(\partial_x z_j + \partial_y z_j \right)
\right]\right\} \; ,
\nonumber \\
{\bar U}_{j,AA}^{(3)} & = & - \sqrt{2} a \left({\bar z}_j \partial_x z_j 
- {\bar z}_j \partial_y z_j\right)
\nonumber \\ & &
+ a^2 \left\{
{\bar z}_j \partial^2_x z_j - 2 {\bar z}_j \partial_{xy} z_j 
+ {\bar z}_j \partial^2_y z_j
+ \sqrt{2} 
\left[ {\bar z}_j \left(\partial_x \zeta_j 
- \partial_y \zeta_j \right) 
+ {\bar \zeta}_j \left(\partial_x z_j - \partial_y z_j \right)
\right]\right\} \; ,
\nonumber \\
{\bar U}_{j,AA}^{(4)} & = & - \sqrt{2} a \left({\bar z}_j \partial_x z_j 
+ {\bar z}_j \partial_y z_j\right)
\nonumber \\ & &
+ a^2 \left\{
{\bar z}_j \partial^2_x z_j + 2 {\bar z}_j \partial_{xy} z_j 
+ {\bar z}_j \partial^2_y z_j
+ \sqrt{2} 
\left[ {\bar z}_j \left(\partial_x \zeta_j 
+ \partial_y \zeta_j \right) 
+ {\bar \zeta}_j \left(\partial_x z_j + \partial_y z_j \right)
\right]\right\} \; .
\eea
For $\Phi^{(i)}_{BB}$, we have the same phase factors as for 
$\Phi^{(i)}_{AA}$ but we have to replace 
${\bar U}_{j,AA}^{(i)}$ by ${\bar U}_{j,BB}^{(i)}$ that are as follows:
\bea
{\bar U}_{j,BB}^{(1)} & = & \sqrt{2} a \left(z_j \partial_x {\bar z}_j 
- z_j \partial_y {\bar z}_j\right)
\nonumber \\ & &
+ a^2 \left\{
z_j \partial^2_x {\bar z}_j - 2 z_j \partial_{xy} {\bar z}_j 
+ z_j \partial^2_y {\bar z}_j
+ \sqrt{2} \left[ z_j \left(\partial_x {\bar \zeta}_j 
- \partial_y {\bar \zeta}_j \right) 
+ \zeta_j \left(\partial_x {\bar z}_j - \partial_y {\bar z}_j \right)
\right]\right\} \; , 
\nonumber \\ 
{\bar U}_{j,BB}^{(2)} & = & \sqrt{2} a \left(z_j \partial_x {\bar z}_j 
+ z_j \partial_y {\bar z}_j\right)
\nonumber \\ & &
+ a^2 \left\{
z_j \partial^2_x {\bar z}_j + 2 z_j \partial_{xy} {\bar z}_j 
+ z_j \partial^2_y {\bar z}_j
+ \sqrt{2} \left[ z_j \left(\partial_x {\bar \zeta}_j 
+ \partial_y {\bar \zeta}_j \right) 
+ \zeta_j \left(\partial_x {\bar z}_j + \partial_y {\bar z}_j \right)
\right]\right\} \; ,
\nonumber \\ 
{\bar U}_{j,BB}^{(3)} & = & - \sqrt{2} a \left(z_j \partial_x {\bar z}_j 
- z_j \partial_y {\bar z}_j\right)
\nonumber \\ & &
+ a^2 \left\{
z_j \partial^2_x {\bar z}_j - 2 z_j \partial_{xy} {\bar z}_j 
+ z_j \partial^2_y {\bar z}_j
- \sqrt{2} \left[ z_j \left(\partial_x {\bar \zeta}_j 
- \partial_y {\bar \zeta}_j \right) 
+ \zeta_j \left(\partial_x {\bar z}_j - \partial_y {\bar z}_j \right)
\right]\right\} \; ,
\nonumber \\ 
{\bar U}_{j,BB}^{(4)} & = & - \sqrt{2} a \left(z_j \partial_x {\bar z}_j 
+ z_j \partial_y {\bar z}_j\right)
\nonumber \\ & &
+ a^2 \left\{
z_j \partial^2_x {\bar z}_j + 2 z_j \partial_{xy} {\bar z}_j 
+ z_j \partial^2_y {\bar z}_j
- \sqrt{2} \left[ z_j \left(\partial_x {\bar \zeta}_j 
+ \partial_y {\bar \zeta}_j \right) 
+ \zeta_j \left(\partial_x {\bar z}_j + \partial_y {\bar z}_j \right)
\right]\right\} \; .
\eea

Finally, we list below the contributions from spin interactions dressed with 
fermions. We defined
\bea
\label{Fplus}
{\cal F}^\pm \equiv \frac{1}{2} 
\left[ {\cal F}^{(e)} ({\vec k}_1,{\vec k}_2)
\pm {\cal F}^{(o)} ({\vec k}_1,{\vec k}_2)\right] \; ,
\eea
with 
\bea
{\cal F}^{(e,o)} ({\vec k}_1,{\vec k}_2) & = & 
\frac{4a^2}{N} 
\exp \left[ i \left( {\vec k}_1 - {\vec k}_2 \right) \cdot {\vec x}_{A,B} 
\right]
\sum_j 
\exp \left[ i \left({\vec k}_1 - {\vec k}_2 \right) \cdot {\vec x}_j \right] 
\Upsilon_j
\; ,
\eea
where
\bea
\Upsilon_j = 4\left\{
2 G^{}_j G^*_j + 2 F^{}_{jy} F^*_{jy} + 2 F^{}_{jx} F^*_{jx} 
-\left( F^{}_{jy} F^*_{jx} + F^{}_{jx} F^*_{jy} \right)  
+\sqrt {2} \left[ \left( F^{}_{jy} - F^{}_{jx} \right) 
G^*_j + G^{}_j  \left(F^*_{jy} -  F^*_{jx} \right) \right]
\right\} \; ,
\eea
contains the contributions from ${\vec \Omega}_i \cdot {\vec \Omega}_j$. 

\section{Effective action $S_{\rm eff}^{(\zeta)}$
\label{AfterConstraintZZeta}}
We display here, in order to facilitate a reproduction of our results, the 
different contributions to the part of the action containing the
$\zeta$-fields (\ref{SFermionZeta}) 
after imposing the constraint (\ref{constzzAfterChange}).
We consider first the part containing temporal derivatives in 
(\ref{SFermionZeta})
\bea
\label{TemporalDerivativesZZeta}
{\bar z} \, \partial_\tau \zeta
+ {\bar \zeta} \, \partial_\tau z & = & 
\frac{\zeta_2^* \, z^{}_2}
{\mid z_1 \mid^2} \, z^{}_1 \, \partial_\tau z_1^*
- \frac{z_2^* \, \zeta_2^{}}{\mid z_1 \mid^2} \, 
z_1^* \, \partial_\tau z^{}_1
+ z^*_2 \, \partial_\tau \zeta^{}_2 - z^{}_2 \, \partial_\tau \zeta_2^* 
\; .
\eea
We consider next the interaction part in (\ref{SFermionZeta}). Here we have 
\bea
G & = & -2 z_1 \left( \frac{z_2^2 \, \zeta_2^*}{\mid z_1 \mid^2}
+ \zeta_{2} \right) \; .
\eea
Then, for the different terms entering (\ref{SFermionZeta}) we have,
\bea
G^* G & = & 
4 \left(\frac{\mid z_2 \mid^4+\mid z_1 \mid^4}{\mid z_1 \mid^2}  
\mid \zeta_2 \mid^2
+ z_2^{*2} \, \zeta_2^2 + z_2^2 \, \zeta_2^{*2} \right) \; ,
\nonumber \\
G^* F^{}_\alpha & = & -2 z^*_1 
\left( \frac{z_2^{*2} \, \zeta^{}_2}{\mid z_1 \mid^2}
+ \zeta^*_2 \right) F^{}_\alpha \; ,
\nonumber \\ 
F^*_\alpha G & = & -2 F^*_\alpha z_1 
\left( \frac{z_2^2 \, \zeta_2^*}{\mid z_1 \mid^2}
+ \zeta_{2} \right) \; .
\eea
\newpage


\begin{thebibliography}{10}

\bibitem{science06}
A. Cho, Science {\bf 314},  1072  (2006).

\bibitem{sachdev03}
S. Sachdev, Rev. Mod. Phys. {\bf 75},  913  (2003).

\bibitem{kivelson03}
S. Kivelson {\it et~al.}, Rev. Mod. Phys. {\bf 75},  1201  (2003).

\bibitem{demler04}
E. Demler, W. Hanke, and S.-C. Zhang, Rev. Mod. Phys. {\bf 76},  909  (2004).

\bibitem{lee06}
P.~A. Lee, N. Nagaosa, and X.-G. Wen, Rev. Mod. Phys {\bf 78},  17  (2006).

\bibitem{shraiman88b}
B. Shraiman and E. Siggia, Phys. Rev. Lett. {\bf 61},  467  (1988).

\bibitem{shraiman89}
B. Shraiman and E. Siggia, Phys. Rev. Lett. {\bf 62},  1564  (1989).

\bibitem{shraiman90}
B. Shraiman and E. Siggia, Phys. Rev. B {\bf 42},  2485  (1990).

\bibitem{shraiman92}
B. Shraiman and E. Siggia, Phys. Rev. B {\bf 46},  8305  (1992).

\bibitem{azaria90}
P. Azaria, B. Delamotte, and T. Jolicoeur, Phys. Rev. Lett. {\bf 64},  3175
  (1990).

\bibitem{azaria93}
P. Azaria, B. Delamotte, F. Delcduc, and T. Jolicoeur, Nucl. Phys. B {\bf 408},
   485  (1993).

\bibitem{chubukov94}
A.~V. Chubukov, S. Sachdev, and T. Senthil, Nucl. Phys. B {\bf 426},  601
  (1994).

\bibitem{azaria95}
P. Azaria, P. Lecheminant, and D. Mouhanna, Nucl. Phys. B {\bf 455},  648
  (1995).

\bibitem{chubukov96}
A.~V. Chubukov and O.~A. Starykh, Phys. Rev. B {\bf 53},  R14729  (1996).

\bibitem{matsuda02}
M. Matsuda {\it et~al.}, Phys. Rev. B {\bf 65},  134515  (2002).

\bibitem{bourges00}
P. Bourges {\it et~al.}, Science {\bf 288},  1234  (2000).

\bibitem{mook02}
H. Mook, P. Dai, and F. Dogan, Phys. Rev. Lett. {\bf 88},  097004  (2002).

\bibitem{hayden04}
S. Hayden {\it et~al.}, Nature {\bf 429},  531  (2004).

\bibitem{hinkov04}
V. Hinkov {\it et~al.}, Nature {\bf 430},  650  (2004).

\bibitem{juricic04}
V. Juricic, L. Benfatto, A. Caldeira, and C.~M. Smith, Phys. Rev. Lett. {\bf
  92},  137202  (2004).

\bibitem{sushkov05}
O. Sushkov and V. Kotov, Phys. Rev. Lett. {\bf 94},  097005  (2005).

\bibitem{lindgard05}
P.-A. Lindg\aa{}rd, Phys. Rev. Lett. {\bf 95},  217001  (2005).

\bibitem{juricic06}
V. Juricic, M. da~Silva~Neto, and C.~M. Smith, Phys. Rev. Lett. {\bf 96},
  077004  (2006).

\bibitem{luescher07}
A. L\"uscher, A. Milstein, and O. Sushkov, Phys. Rev. Lett. {\bf 98},  037001
  (2007).

\bibitem{kaempfer05}
F. K\"ampfer, M. Moser, and U.-J. Wiese, Nucl. Phys. B {\bf 729},  317  (2005).

\bibitem{anderson87}
P.~W. Anderson, Science {\bf 235},  1196  (1987).

\bibitem{zhang88}
F. Zhang and T.~M. Rice, Phys. Rev. B {\bf 37},  3759  (1988).

\bibitem{dagotto94}
E. Dagotto, Rev. Mod. Phys. {\bf 66},  763  (1994).

\bibitem{sorella02}
S. Sorella {\it et~al.}, Phys. Rev. Lett. {\bf 88},  117002  (2002).

\bibitem{wiegmann88}
P. Wiegmann, Phys. Rev. Lett. {\bf 60},  821  (1988).

\bibitem{wiegmann89}
P. Wiegmann, Nucl. Phys. B {\bf 323},  311  (1989).

\bibitem{faddeev88}
L. Faddeev and R. Jackiw, Phys. Rev. Lett {\bf 60},  1692  (1988).

\bibitem{foussats99}
A. Foussats, A. Greco, and O. Zandron, Ann. Phys. {\bf 275},  238  (1999).

\bibitem{weinberg05V1}
S. Weinberg, {\em The Quantum Theory of Field} (Cambridge University Press,
  Cambridge, 2005), Vol.~I.

\bibitem{hubbard63}
J. Hubbard, Proc. R. Soc. London {\bf A276},  238  (1963).

\bibitem{hubbard64a}
J. Hubbard, Proc. R. Soc. London {\bf A277},  237  (1964).

\bibitem{hubbard64b}
J. Hubbard, Proc. R. Soc. London {\bf A281},  401  (1964).

\bibitem{scheunert77}
M. Scheunert, W. Nahm, and V. Rittenberg, J. Math. Phys. {\bf 18},  155
  (1977).

\bibitem{fradkin91}
E. Fradkin, {\em Field Theories of condensed matter systems}, {\em Frontiers in
  Physics} (Addison-Wesley Publishing Company, Redwood City, 1991).

\bibitem{gitman90}
D. Gitman and I. Tyutin, {\em Quantization of Fields with Constraints}
  (Springer, Berlin, 1990).

\bibitem{cornwell89V3}
J. Cornwell, {\em Group Theory in Physics} (Academic Press, London, 1989),
  Vol.~III.

\bibitem{senjanovic76}
P. Senjanovic, Ann. Phys. {\bf 100},  227  (1976).

\bibitem{kim98}
C. Kim {\it et~al.}, Phys. Rev. Lett. {\bf 80},  4245  (1998).

\bibitem{kane89}
C.~L. Kane, P.~A. Lee, and N. Read, Phys. Rev. B {\bf 39},  6880  (1989).

\bibitem{martinez91}
G. Martinez and P. Horsch, Phys. Rev. B {\bf 44},  317  (1991).

\bibitem{liu92}
Z. Liu and E. Manousakis, Phys. Rev. B {\bf 45},  2425  (1992).

\bibitem{brunner00b}
M. Brunner, F.~F. Assaad, and A. Muramatsu, Phys. Rev. B {\bf 62},  15480
  (2000).

\bibitem{muramatsu90}
A. Muramatsu and R. Zeyher, Nucl. Phys. B {\bf 346},  387  (1990).

\bibitem{kuebert95}
C. K\"ubert and A. Muramatsu, Europhys. Lett. {\bf 30},  481  (1995).

\bibitem{klee96}
S. Klee and A. Muramatsu, Nucl. Phys. B {\bf 473},  539  (1996).

\end{thebibliography}

\end{document}